\documentclass{article}
\usepackage{svg}
\usepackage{xcolor}
\usepackage{authblk}
\usepackage{amsmath,amsthm,amsfonts,amssymb,amscd}
\usepackage{amssymb,setspace,array,booktabs,threeparttable,rotating,multirow,pdflscape,amsthm,floatpag,xfrac,enumerate}
\usepackage{lastpage}
\usepackage{enumerate}
\usepackage{amsmath}
\usepackage{fancyhdr, floatpag}
\usepackage{mathrsfs}
\usepackage{graphicx}
\usepackage{listings}
\usepackage{hyperref}
\usepackage{inputenc}
\usepackage{longtable}
\usepackage{threeparttable}
\usepackage{threeparttablex} 
\usepackage{geometry} 
\usepackage{verbatim}
\usepackage{enumerate}
\usepackage{subcaption}
\usepackage{longtable}
\usepackage{dirtytalk}
\usepackage{geometry}
\usepackage{amsmath}
\usepackage{natbib}
\usepackage{verbatim}
\usepackage{longtable}
\usepackage{arydshln}
\usepackage{bbm}
\usepackage[capposition=bottom]{floatrow}
\usepackage{pdflscape}
\usepackage{amsmath,amssymb}
\usepackage{amsmath}
\usepackage{float}
\usepackage{doi}
\usepackage{lineno}
\usepackage{tikz}
\usetikzlibrary{calc}
\usetikzlibrary{arrows.meta, positioning, backgrounds}

\newlength\tindent
\newcolumntype{L}[1]{>{\raggedright\let\newline\\\arraybackslash\hspace{0pt}}m{#1}}
\newcolumntype{C}[1]{>{\centering\let\newline\\\arraybackslash\hspace{0pt}}m{#1}}
\newcolumntype{R}[1]{>{\raggedleft\let\newline\\\arraybackslash\hspace{0pt}}m{#1}}

\newcommand{\ATECausalForest}{-0.0667}
\newcommand{\ATELinearModel}{0.0681}
\newcommand{\SECausalForest}{0.0348}
\newcommand{\SELinearModel}{0.051}

\newcommand{\LinearModelFirstStageRSQ}{0.348}
\newcommand{\CausalForestFirstStageRSQ}{0.308}
\newcommand{\LinearModelFStat}{77.45234}

\title{Optimizing Patient Placement in Normal Care Units: An Instrumental Causal Forest Approach Minimizing Mortality}
\author{Johannes Cordier\thanks{\textbf{Acknowledgments:} I am am grateful to Pierre Alquier, Amanda Kowalski, Jacob Dorn, and Jeffrey McCullough for their valuable feedback. I also thank participants at the Joint Asian and European Workshop on Econometrics and Health Economics and the Annual Workshop in Health Econometrics for helpful comments and suggestions.}}
\affil{\small Chair of Health Economics, Policy and Management, School of Medicine, University of St. Gallen}

\date{31.12.2025}
\begin{document}
\maketitle

\noindent 

\begin{abstract}
   Normal care units (NCU) placement affects health outcomes. NCUs in a hospital have different specialisations. There are patients that can potentially stay in multiple different NCUs. On a given day the NCUs are on different utilisation levels, which also affects health outcomes. Our approach uses instrumental variable causal forests, with emergency admission as an instrument, to estimate how the effect of NCU placement varies across patients and utilisation levels. The results show a clear trade-off between specialisation and utilization. Based on these findings, we design a minimax regret placement policy, using frequentist, Balke-Pearl and Manski bounds, that lowers mortality without capacity expansion. The policy reallocates patients according to their individualized average treatment effects, showing that data-driven patient placement can improve outcomes by using existing resources more efficiently.
\end{abstract}
\newpage

\section{Introduction}\label{intro}

Patient placement in normal care units (NCUs) directly affects outcomes, length of stay, and costs \citet{sharma2022effect}. NCUs are standard hospital wards for patients who are stable enough to not require ICU-level monitoring, but still require continuous inpatient care to recover from a surgeries or other medical interventions. High utilization in these NCUs increases mortality \citep{schilling2010comparison, madsen2014high, abir2020association, sprivulis2006association, elsayed2005pressure}, adverse events \citep{Boyle2013probability}, because nursing staff has less time per patient Kuntz2015stress.

Hospitals operate multiple NCUs with different specializations. An internal medicine NCU focuses on non-operative treatments and diagnostics evaluation, pharmaceutical treatments, and monitoring and a surgical NCU focuses on patients that need a surgical intervention including pre- and post-surgery care. Suboptimal placement into these NCUs, reduces care quality and increases the risk adverse  events \citep{handel2018inpatient, lloyd2005practice, stowell2013hospital, alameda2009clinical, lepage2009use}.

Bed capacities \citep{leuchter2025health} and limited number of nurses \citep{who_health_workforce} creates in the allocation of resources. Utilization levels affect mortality \citep{bosque2023association, Wiseh4977}. Traditional placement strategies rely primarily on diagnosis and risk factors, but overlook there is a fundamental trade-off between specialization and utilization.

To analyze this trade-off, we focus on patients whose primary ICD-10 diagnosis falls into ischemic heart disease (I2-I3), cerebrovascular disease (I6-I7), and selected oncology categories (C1-C4, C7). These conditions are frequently placed to both internal medicine and surgical NCUs, and exhibit meaningful mortality risk. This approach avoids heterogeneity from populations whose placement is determined by strict clinical rules (e.g., obstetrics, psychiatry, pediatrics) and ensures that (1) mortality is a relevant outcome, (2) placement is not mechanically predetermined, and (3) counterfactual placements are realistic. Placement matters for these groups because internal medicine NCUs provide disease-management capacity and subspecialty input, while surgical NCUs specialize in perioperative stability and complication response. The optimal ward is not always clinically obvious at admission.

Hospital operations research has long examined queueing, simulation, and optimization models for bed allocation \citep{Boyle2013probability, Kuntz2015stress, Wiseh4977}, but these approaches rarely incorporate patient-level heterogeneity or causal effects. Empirical work documents that high occupancy and out-lying increase mortality \citep{abir2020association, madsen2014high}, yet conventional regression designs face confounding. Recent advances in causal inference and machine learning (instrumental variables, causal forests, and policy learning \citep{wager2018estimation, athey2016, athey2019machine}) allow individualized treatment effects but have not been integrated with operational decision frameworks. We bridge both strands by combining heterogeneous causal effect estimation with operational policy optimization.\\

The study focuses on the following research question:\\
    How can a patient placement policy be designed to minimize mortality while balancing specialization and utilization?\\

We use Swiss inpatient data from 2012 to 2020 from five Swiss university hospitals. We estimate individualized treatment effects (IATEs) of NCU placement using instrumental variable causal forests \citep{wager2018estimation, athey2016}, with exogenous variation from the daily number of emergency admissions serving as an instrument for placement decisions. This approach accounts for selection into NCUs and captures how the benefit of placement varies with diagnosis, comorbidities, and unit busyness. We translate these IATEs into placement rules using policy learning. A greedy empirical welfare maximization policy provides the benchmark assignment, and we then evaluate minimax regret policies that are robust to sampling uncertainty. To quantify robustness under weaker assumptions, we report (1) welfare under frequentist lower-bound confidence sets, (2) Manski worst-case bounds, and (3) Balke-Pearl bounds exploiting the IV structure \citep{manski1990nonparametric, balke1997bounds}. We compare these policies to observed hospital assignments to measure counterfactual welfare gains and reductions in mortality.

This study provides evidence that placement decisions can be improved without increasing capacity. Reallocating patients based on individualized policy scores reduces mortality risk and smooths NCU utilization. Methodologically, the paper applies causal machine learning with robust policy optimization for a setting where treatment effects are heterogeneous and interdependent across individuals due to capacity constraints. We show that hospitals can reduce preventable mortality through assignment changes rather than resource expansion.

\section*{Data and Methods}\label{datamethods}

\section*{Data and Methods}\label{data}
This section describes the data sources, sample construction, and empirical methods used in the analysis, including the identification strategy, heterogeneous treatment effect estimation, bounding approaches, and the policy evaluation framework.

\subsection{Data}
We evaluate the effect of NCU placement (Internal Medicine vs. Surgical) on in-hospital mortality using administrative hospital data from the Swiss Federal Statistical Office's Medical Statistics of Hospitals. This dataset provides detailed information on socio-demographic characteristics, outcomes, treatments coded according to the Swiss surgical classification (CHOP), and diagnoses coded using ICD-10-GM. The dataset spans 2012-2020 and includes 278 hospitals with approximately 13 million cases.
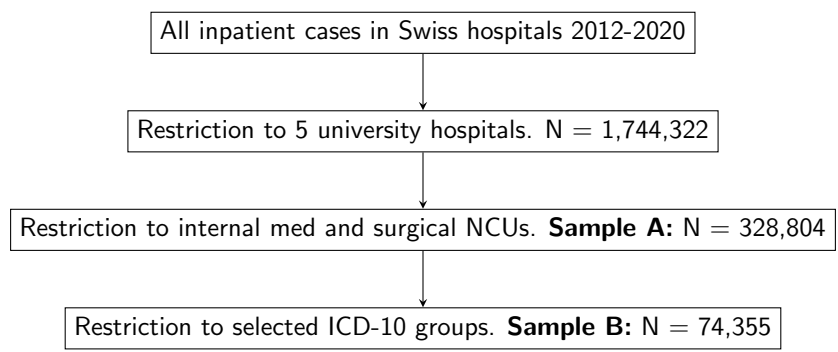
\begin{figure}[ht]
    \centering
    \begin{tikzpicture}[
        node distance=0.75cm,
        >=stealth,
        box/.style={
            rectangle,
            draw,
            minimum width=5cm,
            text centered,
            font=\sffamily
        }
    ]

    \node[box] (start) {All inpatient cases in Swiss hospitals 2012-2020};
    \node[box, below=of start] (univ) {Restriction to 5 university hospitals. N = 1,744,322};
    \node[box, below=of univ] (ncu) {Restriction to internal med and surgical NCUs. \textbf{Sample A:} N = 328,804};
    \node[box, below=of ncu] (icd) {Restriction to selected ICD-10 groups.  \textbf{Sample B:} N = 74,355};

    \draw[->] (start) -- (univ);
    \draw[->] (univ) -- (ncu);
    \draw[->] (ncu) -- (icd);

    \end{tikzpicture}
    \caption{Exclusion flowchart for inpatient hospital sample construction, 2012-2020}
    \label{fig:sample_exclusion_flowchart}
\end{figure}

We use administrative hospital data from the Swiss Federal Statistical Office's Medical Statistics of Hospitals to evaluate the effect of NCU placement (Internal Medicine vs. Surgical) on in-hospital mortality \citep{zellweger2019cause}. This dataset provides detailed information on socio-demographic characteristics, outcomes, treatments coded according to the Swiss surgical classification (CHOP), and diagnoses coded using ICD-10-GM.

We apply several restrictions for our analysis. (1), we focus on the five university hospitals for sample size reasons. (2) to ensure a well-defined choice of NCU placement, we limit the sample to patients admitted either to the internal medicine NCU or the surgical NCU, reducing heterogeneity in admission options. Analyzing more NCUs would drastically reduce the overlap between patients. Because specialties such as gynaecology and obstetrics, psychiatry and psychotherapy, and paediatrics almost exclusively treat women giving birth, psychiatric patients, or children, respectively, these NCUs are not comparable to the general inpatient population.
(3) for clinical comparability, we restrict the sample to patients whose main diagnosis falls within selected ICD-10 groups (I2, I3, I6, I7, C1, C2, C3, C4, and C7). We choose these diagnosis groups because patients with these conditions are treated in both NCUs, ensuring sufficient overlap for identification and they exhibit non-trivial mortality risk. Without an outcome to improve, optimization would neither be clinically relevant or statistically feasible. For example, including orthopedic cases, where mortality is very low and the treatment objective is functional recovery and not survival, would make it nearly impossible to distinguish between patient trajectories, as the outcome offers minimal variation and would therefore lack discriminatory power.

The final sample as shown in Figure \ref{fig:sample_exclusion_flowchart} as Sample B contain 74,355 cases. Sample B contains the cases whose placements are being optimised. Sample A is the is used to estimate the busyness (the number of patients in a NCU on any given day) as a proxy for utilisation.

We use a day-based random train and test split for out-of-sample policy learning \citep{momo2023lengthstaypredictionhospital}. We randomly assign 75\% of hospital-days to the training set and the remaining 25\% to the test set. The causal forest is trained on data from the training set and the policy is built on the test set. The day-based splitting allows us to capture temporal variation in patient characteristics, treatment patterns, and outcomes, while maintaining separation between training and test observations. 

The resulting training set contains 55,766 cases (75\% of 74,355) across 3,195 unique hospital-days, while the test set contains 15,081 cases across 1,065 hospital-days. All reported policy evaluation results, including welfare gains, mortality reductions, and busyness redistributions, are computed exclusively on the  test set. The causal forest estimation, including the training of all nuisance components is build on the training data. The busyness before the placement of patients each day is estimate on the train and test set of Sample A.

\subsubsection*{Dependent Variable}

The primary outcome is in-hospital mortality, defined as death during hospital stay. The dependent variable equals 1 if the patient survived to discharge and 0 if the patient died during the stay. We maximize survival. This outcome is recorded in the administrative data for all cases.

We focus on in-hospital mortality for three reasons: (1) it is a critical and unambiguous outcome that reflects inpatient care quality. (2) it is consistently recorded across all hospitals and time periods, ensuring data quality. (3) for the included cardiovascular and cancer diagnoses, in-hospital mortality captures acute complications and treatment failures. 

\subsubsection*{Independent Variable}

All 74,355 cases in the sample have observed NCU placements recorded in the administrative data through the primary cost center variable. The treatment assignment $D_i$ indicates whether patient $i$ was assigned to the internal medicine NCU ($D_i = 1$) or the surgical NCU ($D_i = 0$) during their hospital stay. These observed assignments reflect real-world placement decisions made by hospital staff under operational constraints and clinical judgment. These two types of NCUs differ fundamentally in their clinical focus, staffing, workflows, and available resources, creating distinct care environments that may differentially affect patient outcomes.

Internal medicine NCUs generally treat patients with acute or chronic non-surgical illnesses, including cardiac, respiratory, infectious, metabolic, renal, and neurological conditions \citep{kc2009, aiken2002}. Surgical NCUs treat patients undergoing surgery, focusing on postoperative recovery and complication management \citep{needleman2011}. 

The primary difference lies in care focus: internal medicine NCUs emphasize disease management, diagnostic evaluation, and subspecialist coordination, while surgical NCUs follow standardized postoperative care pathways.

\subsubsection*{Control Variables}

The causal forest uses 2,122 variables to capture patient heterogeneity and estimate conditional treatment effects:

\begin{itemize}
    \item Diagnoses (1,209 variables): ICD-10 diagnoses are grouped to the three-digit level and encoded as binary indicators. Only main diagnoses or listed in the Elixhauser comorbidity index are included, as these represent pre-existing conditions known at the time of placement decision. \citep{elixhauser1998comorbidity}.

    \item Treatments (890 variables): CHOP are grouped to the three-digit level and encoded as binary indicators. Only the main treatment is recorded, as these treatments are known at the time of placement decision. 
    
    \item Demographic variables (4 variables): Age (continuous), sex (binary), Swiss nationality (binary), and admission type (emergency vs. elective, binary). Age is included as a continuous variable.
    
    \item Temporal variables (12 variables): Day-of-week indicators (7 binary variables), month-of-year indicators (12 binary variables, though only 11 are included after removing one reference category), and year indicators (9 binary variables for years 2012-2020). These capture seasonal patterns, weekday effects, and temporal trends in patient populations and hospital practices.
    
    \item Hospital indicators (5 variables): Binary indicators for each of the five university hospitals, controlling for unobserved hospital-specific factors such as patient populations, referral patterns, and organizational practices.
    
    \item Busyness measures (2 variables): Current occupancy levels in the internal medicine NCU and surgical NCU, measured as the number of patients currently admitted to each unit on the day of the index patient's admission.
\end{itemize}

\subsubsection*{Instrumental Variable}

To address potential selection bias in NCU assignment, we employ an instrumental variable based on the daily number of emergency admissions to the hospital. We construct the instrument as a binary indicator equal to one if the number of emergency cases admitted on the same day as the index patient (explicitly excluding the index case itself if it is classified as an emergency) exceeds the hospital-year specific median, and zero otherwise. Days with above-median emergency admissions generate exogenous variation in NCU busyness and affect the likelihood of a patient being assigned to a particular NCU, independent of individual patient characteristics or anticipated outcomes.

\subsection{Methods}
This section describes the methodological framework used to estimate treatment effects and evaluate placement policies for normal care units. The analysis combines an instrumental variable design with causal forests to estimate heterogeneous treatment effects at the patient level, using daily emergency admission volume as a source of exogenous variation in NCU assignment. Conditional effects are examined along clinical specialisation and utilisation dimensions to capture how diagnostic requirements and occupancy levels influence outcomes. Bound estimation provides identification regions under varying assumptions, and these results inform the subsequent policy evaluation through a minimax regret framework that compares observed assignments with counterfactual placement strategies. Together, these components provide a coherent approach for estimating effects and assessing the implications for decision making under operational and clinical constraints.

\subsubsection{Causal Forests and Instrumental Variables}

\subsubsection{Setup and Identification}
In the first step, we employ an instrumental variable causal forest to estimate the heterogeneous treatment effects of NCU admission on patient outcomes. This approach accounts for potential confounding by exploiting exogenous variation in the daily number of emergency admissions, which serves as the instrument. The patient-specific policy scores from this estimation quantify the expected benefit for each patient when admitted to a less congested but still appropriately specialized NCU.

The estimation relies on the instrumental variable framework, where we construct the instrument from the number of emergency admissions (excluding the index case if classified as emergency). Fluctuations in emergency admissions affect the probability of admission to a particular unit but do not directly influence patient outcomes except through this assignment, making the instrument plausibly exogenous.

A key assumption underlying the causal inference framework is the Stable Unit Treatment Value Assumption (SUTVA), which requires that the potential outcomes for each patient are unaffected by the treatment assignment of other patients \citep{rubin1974estimating, athey2019machine}. In the context of NCU placement, SUTVA could be violated if the assignment of one patient to a particular unit affects the outcomes of other patients through mechanisms such as increased congestion, reduced staff attention, or resource competition. While this study controls for NCU busyness levels by incorporating them as covariates in the causal forest estimation, these controls provide only a partial solution. The busyness measure captures the overall patient load but may not fully account for dynamic interactions between patient assignments or unmeasured aspects of care quality under varying workload conditions. Therefore, while the instrumental variable approach and the inclusion of busyness controls strengthen causal identification, potential violations of SUTVA remain a conceptual limitation of the analysis, and the estimated treatment effects should be interpreted as conditional on the observed levels of NCU utilization.

The causal forest partitions the data into subgroups with similar covariate profiles and estimates treatment effects flexibly within these groups. This accommodates non-linearities and complex interactions between patient characteristics, which is often not possible with standard linear IV models. We estimate all components, including the conditional mean outcomes, propensity scores, and instrument probabilities, through the causal forest procedure.

While we employ causal forests in this study, alternative nonparametric estimation methods such as kernel regression, support vector machines, k-nearest neighbors, or neural networks could potentially be used for estimating the conditional mean outcomes and propensity scores, though Random Forests have demonstrated strong empirical performance and theoretical consistency for causal inference applications \citep{breiman2001random, biau2012analysis, scornet2015consistency, wager2018estimation}. We choose causal forests for their theoretical foundation for heterogeneous treatment effect estimation and their ability to provide consistent estimates while maintaining computational efficiency for moderately large datasets \citep{wager2018estimation}.

In summary, by leveraging exogenous variation in emergency admissions, the methodology provides patient-specific treatment effect estimates that support optimal patient assignment decisions while ensuring unbiased causal inference. The instrument strength is moderate, and causal identification depends on the validity of the instrument assumptions discussed in the identification section.

\subsubsection*{Setup and Identification for Instrumental Variables}

To estimate the heterogeneous treatment effects of NCU admission on patient outcomes while addressing potential confounding, we employ an instrumental variable (IV) causal forest approach. The IV method exploits exogenous variation in the daily number of emergency admissions, enabling robust causal inference in the presence of selection bias common in observational hospital data.

Let $Y_i$ denote the outcome for patient $i$, $D_i$ the treatment indicator (admission to a specific NCU), and $Z_i$ an instrumental variable that influences $D_i$ but does not directly affect $Y_i$. The standard potential outcomes framework defines the individual treatment effect as \citep{rubin1974estimating}:
\[
\tau_i = Y_i(1) - Y_i(0),
\]
where $Y_i(1)$ and $Y_i(0)$ are the potential outcomes under treatment and control, respectively. However, in the instrumental variable setting, the local average treatment effect (LATE) is estimated rather than the average treatment effect (ATE), and this is conditional on compliance with the instrument.

For $Z_i$ to be a valid instrumental variable, several identifying assumptions must be satisfied \citep{angrist2009mostly}:

\begin{enumerate}
    \item \textbf{Instrument Relevance:} The instrument $Z_i$ must be correlated with the treatment $D_i$, meaning that the number of emergency admissions captured by $Z$ should influence the likelihood of a patient being assigned to a specific NCU. Mathematically, this is expressed as:
    \[
    \mathbb{E}[D_i \mid Z_i=1] \neq \mathbb{E}[D_i \mid Z_i=0].
    \]
    In this context, higher emergency admissions on a given day affect the probability of a patient being assigned to one unit over another. While the instrument strength is moderate, controlling for patient and hospital characteristics helps ensure causal identification.

    \item \textbf{Instrument Exogeneity:} Conditional on the observed covariates $X_i$, the instrument $Z_i$ must be independent of the potential outcomes $Y_i(1)$ and $Y_i(0)$. This ensures that $Z$ does not directly affect the outcome except through its influence on the treatment $D_i$. Formally:
    \[
    Y_i(1), Y_i(0) \perp Z_i \mid X_i.
    \]
    This assumption is plausible because we construct $Z$ based on exogenous variation in emergency admissions, which is unrelated to individual patient characteristics. The exclusion of the index case from the emergency count further reinforces exogeneity.

    \item \textbf{Exclusion Restriction:} The instrument $Z_i$ affects the outcome $Y_i$ only through its effect on the treatment $D_i$. This implies that any impact of $Z$ on patient outcomes is mediated entirely by assignment to a specific NCU. Formally:
    \[
    \mathbb{E}[Y_i \mid Z_i, X_i] = \mathbb{E}[Y_i \mid D_i, X_i].
    \]
    In this setting, the number of emergency admissions captured by $Z$ influences patient outcomes only through its effect on NCU assignment.

    \item \textbf{Monotonicity:} The effect of the instrument on treatment assignment must be monotonic. This means that for all patients, an increase in $Z$ (i.e., more emergency admissions) either increases or does not change the likelihood of being assigned to a particular NCU. Formally:
    \[
    D_i(Z_i=1) \geq D_i(Z_i=0) \quad \forall i.
    \]
    This assumption ensures that there are no defiers (patients who would move in the opposite direction of the instrument's effect). As in every observational study, some defiers exist, as shown in Fig. \ref{fig:compliance_score}, where negative compliance score values appear. However, their presence is not substantial enough to introduce computational issues.
\end{enumerate}

These assumptions together ensure that \( Z \) is a valid instrument, allowing us to estimate the causal effect of NCU assignment on patient outcomes. By leveraging exogenous variation in the daily number of emergency admissions, \( Z \) provides a robust basis for addressing potential confounding and selection bias in the analysis. This approach follows established practices in health economics, where emergency admission patterns have been successfully used as instrumental variables in healthcare settings \citep{moler2022local, hutchings2022effectiveness}. The use of emergency admissions as an instrument has demonstrated validity for estimating causal effects in hospital settings, particularly when analyzing surgical decisions and resource allocation \citep{cooper2022higher}. However, as in most observational studies, the validity of these assumptions cannot be fully tested, and some residual confounding may remain.

\subsubsection{IATE Estimation from Instrumental Causal Forest}

The estimation of individualized average treatment effects (IATEs) builds upon the augmented inverse-propensity weighted framework from \citet{robins1994estimation}, integrated with instrumental variable causal forests as developed by \citet{wager2018estimation, grfpackage, Chernozhukov2018dml}, extending the random forest methodology of \citet{breiman2001random}. Implementation relies on the \texttt{grf} package in R.

Following \citet{lechner2024cmcf}, individualized local instrumental causal effects can be formalized as the ratio of outcome differences to treatment differences across instrument values, conditional on patient characteristics:
\begin{equation}
\text{IATE}_{IV}(x) = \frac{\mathbb{E}[Y \mid Z=1, X=x] - \mathbb{E}[Y \mid Z=0, X=x]}{\mathbb{E}[W \mid Z=1, X=x] - \mathbb{E}[W \mid Z=0, X=x]}
\label{eq:iate_iv}
\end{equation}

Within a single instrumental causal tree, local IV effects emerge as ratios of outcome and treatment differences between instrument groups in each terminal leaf \citep{athey2016}. For tree $b$ constructed from sample $S$:
\begin{equation}
\hat{\tau}_{IV}(x) = 
\frac{
\frac{1}{N_l^{Z=1}} \sum_{i: X_i \in l(x;b), Z_i = 1} Y_i - 
\frac{1}{N_l^{Z=0}} \sum_{i: X_i \in l(x;b), Z_i = 0} Y_i
}{
\frac{1}{N_l^{Z=1}} \sum_{i: X_i \in l(x;b), Z_i = 1} W_i - 
\frac{1}{N_l^{Z=0}} \sum_{i: X_i \in l(x;b), Z_i = 0} W_i
}
\label{eq:iv_tree}
\end{equation}
where $l(x;b)$ represents leaf $l$ in tree $b$ containing $x$, and $N_l^{Z=1}$, $N_l^{Z=0}$ denote observation counts for each instrument value within the leaf.

An instrumental causal forest aggregates predictions from $B$ individual IV trees \citep{wager2018estimation}. Each tree produces a local IV effect $\hat{\tau}_{IV,b}(x)$, combined via simple averaging:
\begin{equation}
\hat{\tau}_{IV}(x) = \frac{1}{B} \sum_{b=1}^{B} \hat{\tau}_{IV,b}(x)
\end{equation}

The tree-growing procedure adapts random forest splitting principles \citep{breiman2001random}, but diverges by maximizing local IV effect heterogeneity across leaves rather than prediction accuracy. Consistency and asymptotic normality require "honest" estimation \citep{wager2018estimation}, whereby observations serve exclusively for either split placement or effect estimation, never both.

Honesty implementation follows a double-sample procedure: training data divides into subsamples $I$ and $J$ for each tree. Subsample $J$ determines splits that maximize local IV effect variance, while subsample $I$ provides the estimation sample. Randomization across trees ensures each observation participates in both roles across the forest.

Rather than direct leaf-level aggregation, the forest constructs adaptive neighborhoods through data-driven weights $\alpha(x,i)$ \citep{athey2019machine}. For test point $x$, training observation $i$ receives weight proportional to co-occurrence frequency across tree leaves:
\begin{equation}
\alpha(x,i) = \frac{1}{B} \sum_{b=1}^{B} \frac{\mathbf{1}\{X_i \in l_b(x)\}}{|l_b(x)|}
\label{eq:forest_weights_iv}
\end{equation}

High-dimensional confounding robustness and valid inference require orthogonalized pseudo-outcomes. With nuisance functions $\hat{m}(X_i) = \mathbb{E}[Y_i \mid X_i]$ and $\hat{e}(X_i) = \mathbb{E}[Z_i \mid X_i]$, the orthogonalized IV pseudo-outcome becomes:
\begin{equation}
\tilde{Y}_i^{IV} = \frac{(Y_i - \hat{m}(X_i))(Z_i - \hat{e}(X_i))}{\hat{\tau}_W(X_i)}, \quad
\hat{\tau}_W(X_i) = \mathbb{E}[W_i \mid Z=1, X_i] - \mathbb{E}[W_i \mid Z=0, X_i]
\label{eq:pseudo_outcome_iv}
\end{equation}

The IV IATE estimate emerges as a weighted average of orthogonalized pseudo-outcomes:
\begin{equation}
\hat{\tau}_{IV}(x) = \sum_{i=1}^{n} \alpha(x,i) \tilde{Y}_i^{IV}
\label{eq:grf_iate_iv}
\end{equation}

The instrumental variable causal forest is implemented using grf package in R \citep{grfpackage}. The model is estimated with standard hyperparameter settings recommended by the authors.

\subsubsection{Bounding Approaches for Individualized Average Treatment Effects}

For each patient only the realized outcome under the actual treatment is observed, while the counterfactual remains unobserved \citep{holland1986statistics}. This missing information introduces uncertainty in estimating treatment effects, which is particularly critical when decisions can affect mortality. To address this, we estimate three lower and upper bounds: (1) frequentist bounds, where uncertainty arises from sampling variability, are essentially the confidence intervals. (2) Balke-Pearl bounds exploit the instrumental variable to tighten the bounds \citep{balke1997bounds}. (3) Manski worst-case bounds have minimal assumptions \citep{manski1990nonparametric}. Ultimately we want to use these bounds for decision making and discrimination between patients. This requires that all bounds are conditioned on the observed covariates $X_i$, so that bounds are defined at the patient level and not at the sample average.

\subsubsection*{Frequentist Confidence Interval Bounds}

The frequentist approach quantifies uncertainty arising from sampling variability in treatment effect estimation. For each patient, I obtain a point estimate of the treatment effect $\hat{\tau}(X_i)$ and its estimated variance $\hat{\sigma}^2(X_i)$ using causal forests \citep{wager2018estimation}. Confidence intervals are defined conditional on $X_i$ at the 95\% confidence level ($\alpha = 0.05$):
\begin{equation}
    \tau(X_i) \in 
    \left[
    \hat{\tau}(X_i) - z_{\alpha/2}\sqrt{\hat{\sigma}^2(X_i)},\,
    \hat{\tau}(X_i) + z_{\alpha/2}\sqrt{\hat{\sigma}^2(X_i)}
    \right].
\end{equation}
These intervals are therefore individualized confidence sets that adjust to the local data distribution surrounding each patient. Conditioning on $X_i$ ensures that the resulting bounds reflect patient specific uncertainty rather than a population wide interval.

\subsubsection*{Balke-Pearl Instrumental Variable Bounds}

Balke-Pearl bounds \citep{balke2022probabilistic,balke1994counterfactual,balke1997bounds} exploit instrumental variable assumptions to tighten the admissible set of treatment effects relative to worst case scenarios. The approach leverages the IV structure to restrict the joint distribution of potential outcomes beyond what observed data alone permit. The bounds are constructed conditional on $X_i$ and require estimating the conditional probabilities of outcome $Y$, treatment $W$, and instrument value $Z$ given covariates.

Let $p_{ywz}(X_i)$ denote the conditional probability $\mathbb{P}(Y_i = Y, D_i = W, Z_i = Z \mid X_i)$ for $Y, D, Z \in \{0,1\}$. The individualized Balke-Pearl bounds for the treatment effect are:
\begin{align}
\tau_{\text{lower}}(X_i) &= \max\Big\{
    -1 + p_{100}(X_i) + p_{110}(X_i) + p_{101}(X_i) + p_{111}(X_i), \nonumber \\
    &\qquad -1 + p_{001}(X_i) + p_{011}(X_i) + p_{101}(X_i) + p_{111}(X_i), \nonumber \\
    &\qquad -1 + p_{010}(X_i) + p_{011}(X_i) + p_{110}(X_i) + p_{111}(X_i)
\Big\}, \\
\tau_{\text{upper}}(X_i) &= \min\Big\{
    1 - p_{000}(X_i) - p_{010}(X_i) - p_{001}(X_i) - p_{011}(X_i), \nonumber \\
    &\qquad 1 - p_{000}(X_i) - p_{010}(X_i) - p_{101}(X_i) - p_{111}(X_i), \nonumber \\
    &\qquad 1 - p_{100}(X_i) - p_{110}(X_i) - p_{001}(X_i) - p_{011}(X_i), \nonumber \\
    &\qquad 1 - p_{100}(X_i) - p_{110}(X_i) - p_{101}(X_i) - p_{111}(X_i)
\Big\}.
\end{align}

All eight conditional probabilities $p_{YDZ}(X_i)$ are estimated using the causal forest structure. For each observation $i$, the forest identifies similar training observations based on covariate proximity, and these local neighborhoods are used to estimate the joint probabilities conditioning on $X_i$. Since both the probabilities and the resulting bounds are conditioned on $X_i$, the Balke-Pearl region is individualized and reflects local covariate structure. The IV-based restrictions make Pearl bounds tighter than Manski bounds while remaining more conservative than frequentist confidence intervals.

\subsubsection*{Manski Worst Case Bounds}

Manski bounds \citep{manski1990nonparametric, manski2003partial} provide the widest admissible range for treatment effects under minimal assumptions. The approach requires only the support of the potential outcomes and conditions all quantities on the observed covariates $X_i$. Let $W_i\in\{0,1\}$ denote treatment assignment and $Y_i\in\{0,1\}$ the binary outcome. Define
\[
p_1(X_i)=\mathbb{P}(W_i=1\mid X_i),\qquad
p_0(X_i)=\mathbb{P}(W_i=0\mid X_i),
\]
and the observed conditional means
\[
\mu_1(X_i)=\mathbb{E}[Y_i\mid W_i=1,X_i],\qquad
\mu_0(X_i)=\mathbb{E}[Y_i\mid W_i=0,X_i].
\]

The bounds for the potential outcome distributions conditional on $X_i$ are
\begin{align}
\mathbb{E}[Y_i(1)\mid X_i]_{\text{lower}} &= \mu_1(X_i)\,p_1(X_i) + 0\cdot p_0(X_i),\\
\mathbb{E}[Y_i(1)\mid X_i]_{\text{upper}} &= \mu_1(X_i)\,p_1(X_i) + 1\cdot p_0(X_i),\\[0.4em]
\mathbb{E}[Y_i(0)\mid X_i]_{\text{lower}} &= \mu_0(X_i)\,p_0(X_i) + 0\cdot p_1(X_i),\\
\mathbb{E}[Y_i(0)\mid X_i]_{\text{upper}} &= \mu_0(X_i)\,p_0(X_i) + 1\cdot p_1(X_i).
\end{align}

The individualized Manski bounds for the treatment effect are
\begin{equation}
    \tau(X_i) \in 
    \left[
    \mathbb{E}[Y_i(1)\mid X_i]_{\text{lower}}
-
\mathbb{E}[Y_i(0)\mid X_i]_{\text{upper}},\,
    \mathbb{E}[Y_i(1)\mid X_i]_{\text{upper}}
-
\mathbb{E}[Y_i(0)\mid X_i]_{\text{lower}}
    \right].
\end{equation}

Since every probability and conditional expectation is indexed by $X_i$, the resulting interval is an individualized bound rather than a population wide region. These bounds represent the maximum uncertainty consistent with the observed data and binary support of the outcome. No restrictions on selection, monotonicity, or exclusion are imposed, which makes Manski bounds the most conservative of the bounding approaches used in this analysis.

\subsubsection{Policy Framework}

In my setting, the average treatment effect does not provide any useful interpretation because the treatment effect is not constant across the patients. The treatment effect is heterogeneous and is depending on the patient characteristics, the NCU busyness, and the NCU specialization. Additionally, not all patients can be placed into one NCU, because of bed capacity constraints. Therefore, a policy that is optimising NCU placement is needed.

Treatment effects vary across patients and depend on clinical characteristics, NCU specialisation, and utilisation conditions. Bed capacity prevents assigning all patients to the same NCU, so placement must be evaluated as an allocation problem rather than as a marginal treatment problem. The policy framework is therefore based on the assignment of $P$ patients per day to two NCUs with the objective of improving survival while respecting capacity. The policy is evaluated through welfare and regret, where welfare measures expected outcomes under a placement rule and regret measures how much welfare is lost relative to the best feasible alternative.

A configuration $\pmb{\pi} = (\pi_1, \ldots, \pi_P)$ is a vector where $\pi_i \in \{0,1\}$ indicates the NCU assignment for patient $i$ (internal medicine if $\pi_i = 1$, surgical if $\pi_i = 0$). Let $D_i^{\text{obs}}$ denote the observed assignment for patient $i$. The unrestricted feasible set is
\begin{equation}
    \pmb{\Pi}^{\text{full}} = \{\pmb{\pi} \in \{0,1\}^P\},
\end{equation}
containing all $2^P$ possible assignment configurations.

Two constraints restrict the feasible set. First, the capacity constraint ensures that NCU capacity limits are respected. Let $C_{h,y,1}$ and $C_{h,y,0}$ denote the maximum bed capacity for internal medicine and surgical NCUs in hospital $h$ and year $y$, respectively. The capacity-constrained feasible set is
\begin{equation}
    \pmb{\Pi}^{\text{cap}} = \left\{\pmb{\pi} \in \pmb{\Pi}^{\text{full}} : \sum_{i=1}^{P} \pi_i \leq C_{h,y,1}, \quad \sum_{i=1}^{P} (1-\pi_i) \leq C_{h,y,0}\right\}.
\end{equation}

Second, the \textbf{reassignment constraint} limits operational disruption by bounding how many patients can be moved from observed placements. For a reassignment limit $\rho \in [0,1]$ (e.g., $\rho = 0.10$ for 10\%), the reassignment-constrained feasible set is
\begin{equation}
    \pmb{\Pi}^{\rho} = \left\{\pmb{\pi} \in \pmb{\Pi}^{\text{cap}} : \sum_{i=1}^{P} \mathbbm{1}_{\{\pi_i \neq D_i^{\text{obs}}\}} \leq \lfloor \rho \cdot P \rfloor\right\}.
\end{equation}

The capacity constraint differs fundamentally from the reassignment constraint: capacity limits reflect physical infrastructure (available beds), while reassignment limits reflect operational feasibility (the number of placement changes hospitals can practically implement on a given day). Section \ref{sec:complexity} describes $\pmb{\Pi}$ and its effect on computational feasibility.

\subsubsection*{Welfare Definition}

Welfare is defined as the expected sum of survival gains (treatment effects) under a given assignment configuration $\pmb{\pi} \in \pmb{\Pi}$. Let $\tau_i(\pmb{\pi})$ denote the individualized treatment effect (change in survival probability) for patient $i$ under configuration $\pmb{\pi}$. The welfare of configuration $\pmb{\pi}$ is
\begin{equation}
    W(\pmb{\pi}) = \mathbb{E}\Bigg[\sum_{i=1}^{P} \tau_i(\pmb{\pi})\Bigg].
\end{equation}

Since $\tau_i(\pmb{\pi})$ is not directly observable, welfare is estimated using predicted treatment effects from the instrumental variable causal forest. The empirical welfare for configuration $\pmb{\pi}$ is
\begin{equation}
    \hat{W}(\pmb{\pi}) = \sum_{i=1}^{P} \hat{\tau}_i(\pmb{\pi}),
\end{equation}
where $\hat{\tau}_i(\pmb{\pi})$ represents the estimated IATE for patient $i$ when assigned according to $\pmb{\pi}$, accounting for the busyness levels in both NCUs that result from the full configuration. These treatment effects incorporate both specialization benefits (clinical match between patient and NCU) and utilization costs (impact of NCU busyness on care quality).

\subsubsection*{Regret Definition}
Building on the growing literature on optimal treatment assignment, initiated by \citet{manski2004statistical} and advanced by \citet{hirano2009asymptotics, stoye2009minimax, kitagawa2018should}, we apply the minimax regret framework to settings where placement decisions jointly affect multiple patients simultaneously. The minimax regret criterion is particularly well suited to healthcare applications, where decision-makers must account for uncertainty in treatment effects and safeguard against harmful assignment errors \citep{manski2007adaptive, pourrezaiekhaligh2023minimax}.

Regret compares the welfare of a given assignment to the best achievable assignment in the feasible set. For configuration $\pmb{\pi}$, the regret is
\begin{equation}
    \widehat{\text{Regret}}(\pmb{\pi})
    = \max_{\pmb{\pi}'\in\pmb{\Pi}}
    \hat{W}(\pmb{\pi}')
    - \hat{W}(\pmb{\pi}).
\end{equation}

\subsubsection*{Welfare Maximization Policy}

The welfare maximization policy selects the assignment configuration that maximizes expected survival gains based on point estimates from the causal forest:
\begin{equation}
    \pmb{\pi}^{WM}
    =
    \arg\max_{\pmb{\pi}\in\pmb{\Pi}}
    \hat{W}(\pmb{\pi}).
\end{equation}

This greedy empirical welfare maximization rule serves as the baseline comparison policy. It prioritizes configurations with the highest predicted welfare without explicit consideration of estimation uncertainty. The policy is implemented either through exact enumeration for small cohorts ($n \leq 22$) or through a greedy approximation that ranks patients by their estimated treatment effects and assigns them to NCUs while respecting capacity constraints.

\subsubsection*{Minimax Regret Policies}

Because welfare is estimated and treatment effects are uncertain, regret is also evaluated with lower bound welfare. This replaces point estimates with welfare computed from the lower bounds of the individualized treatment effect intervals introduced in the bounding section. For each approach (Frequentist, Manski, Pearl), lower bound welfare is written as $W^{\text{lower}}(\pmb{\pi})$ and is constructed by aggregating the lower bound IATEs across patients. Regret under uncertainty is then
\begin{equation}
\text{Regret}(\pmb{\pi})
=
\max\Bigg\{
\max_{\pmb{\pi}'\in\pmb{\Pi}}
\hat{W}(\pmb{\pi}')
-
\hat{W}(\pmb{\pi}),
\;
\max_{\pmb{\pi}'\in\pmb{\Pi}}
W^{\text{lower}}(\pmb{\pi}')
-
W^{\text{lower}}(\pmb{\pi})
\Bigg\}.
\end{equation}

The minimax regret rule selects the assignment that performs best under the worst regret scenario:
\begin{equation}
    \pmb{\pi}^{MR}
    =
    \arg\min_{\pmb{\pi}\in\pmb{\Pi}}
    \text{Regret}(\pmb{\pi}).
\end{equation}

Three minimax regret policies are considered, each based on a different lower bound construction: $\pmb{\pi}^{MR}_{F}$, $\pmb{\pi}^{MR}_{M}$, and $\pmb{\pi}^{MR}_{P}$, where $F$ uses frequentist lower bounds, $M$ uses Manski worst case bounds, and $P$ uses Pearl bounds. These robust policies hedge against estimation uncertainty by selecting assignments that minimize worst-case regret across the admissible range of treatment effects. The policies are compared with the welfare maximization policy $\pmb{\pi}^{WM}$ and with the observed hospital placement $\pmb{\pi}^{obs}$.

The policy operates in the following steps:

\begin{enumerate}
    \item \textbf{Daily Patient Assignment:} Each test day, patients arrive and must be assigned to one of the two NCUs.
    \item \textbf{Outcome Prediction:} Use the instrumental variable causal forest to predict individual average treatment effects (IATEs) for each patient under (all) possible assignments.
    \item \textbf{Welfare Calculation:} Compute estimated welfare for each possible assignment configuration (or a subset following the greedy approximation) using the predicted outcomes from the causal forest.
    \item \textbf{Regret Minimization:} Calculate regret for each possible assignment (or a subset following the greedy approximation) by comparing predicted welfare against the optimal feasible assignment, following the minimax regret criterion, using the three different bounding methodologies.
    \item \textbf{Decision Rule:} The optimal assignment is determined by the optimal policy score. In the welfare maximization approach the maximum welfare and minimax approaches the minimum regret, using the three different bounding methodologies.
\end{enumerate}

\subsubsection{Computational Complexity}\label{sec:complexity}

For $n$ patients per hospital-day, optimal policy selection requires evaluating all possible assignment configurations:

\begin{equation}
    |\pmb{\Pi}| = 2^n
\end{equation}

Daily cohorts range from 1 to 59 patients. The median cohort generates $2^{29} \approx 537$ million configurations. The maximum generates $2^{59} \approx 5.76 \times 10^{17}$. Exact estimation becomes infeasible beyond moderate cohort sizes.

The computations are further constrained by NCU capacity limits. Each NCU has a fixed number of beds, defined by year-hospital specific capacity constraints. These physical constraints reduce the feasible configuration space substantially, as assignments that would exceed capacity are not admissible. This capacity constraint differs from the reassignment limit (which restricts how many patients can be moved from observed placements): the capacity constraint reflects the actual number of beds available in each unit.

Three factors maintain computational feasibility in certain settings. We require $(n-1)n$ IATE estimates, as for each patient we examine every possible distribution of the other patients across NCUs. (2) $31.6\%$ of hospital-days have $n \leq 22$ (at most $2^{22} \approx 4.19$ million configurations), which is manageable. (3) reassignment limits can reduce the feasible space dramatically: a 10\% reassignment limit on a 30-patient cohort with equal number if patients in both NCUs shrinks $2^{29}$ to approximately $\sum_{j=k-3}^{k+3} \binom{29}{j}$, cutting several orders of magnitude. 

We adopt a hybrid strategy: exact enumeration for $n \leq 22$ (451 of 1,427 test days, 31.6\%), greedy approximation for $n > 22$ (976 days, 68.4\%). Greedy algorithms rank patients by policy-specific scores (point estimates for welfare maximization, averaged with bounds for minimax regret variants) and select the top $j$ for internal medicine at each valid busyness level $j$ that respects the year-hospital specific capacity constraints. The number of required estimations decreases from $O(2^n)$ to $O(4n \log n)$. 

\section{Results}

This section presents the empirical findings in four parts. (1) We establish identification and instrument validity through common support analysis and first-stage results. (2) We report average and conditional treatment effects, showing that there is evidence for a specialization and utilization effect. (3) We present the policy results, comparing observed placement decisions with the data-driven policy in terms of welfare gains, patient reassignments, and operational feasibility. (4) We examine hospital-specific heterogeneity and discuss the broader implications for implementation.

\subsection{Descriptive Statistics}

\begin{table}[ht]
\centering
\begin{tabular}{lrrr}
  \hline
Metric & Overall & Int.Med & Surg. \\ 
  \hline
N & 74355 & 44726 & 29629 \\ 
  Survival & 0.97 & 0.96 & 0.98 \\ 
  Int.Med.(W) & 0.60 & 1.00 & 0.00 \\ 
  High Emergency (Z) & 0.46 & 0.45 & 0.48 \\ 
   \hline
\end{tabular}
\caption{Summary statistics for N, Y, W, and Z.} 
\label{tab:summary_statistics}
\end{table}

Table \ref{tab:summary_statistics} provides an overview of patient characteristics and NCU placements in the analysis sample of 74,355 cases. Approximately 60\% of patients are placed in internal medicine NCUs and 40\% in surgical NCUs. The overall survival rate is 96\%, with an in-hospital mortality rate of 4\% showing minimal variation between NCU types at the aggregate level.

The distribution of patients across diagnosis groups and NCUs reveals considerable variation that reflects both clinical specialization and capacity constraints. Cardiovascular diagnoses (I2, I3, I6, I7) are predominantly managed in internal medicine NCUs, while oncological diagnoses (C1-C4, C7) show greater variation, with some cancer types more frequently assigned to surgical NCUs. Hospital-level analysis reveals substantial heterogeneity in placement patterns, indicating that local factors, including organizational structure, referral patterns, and resource availability, significantly influence NCU assignment decisions.

\begin{figure}[htp]
    \centering
    \includegraphics[width=0.9\textwidth]{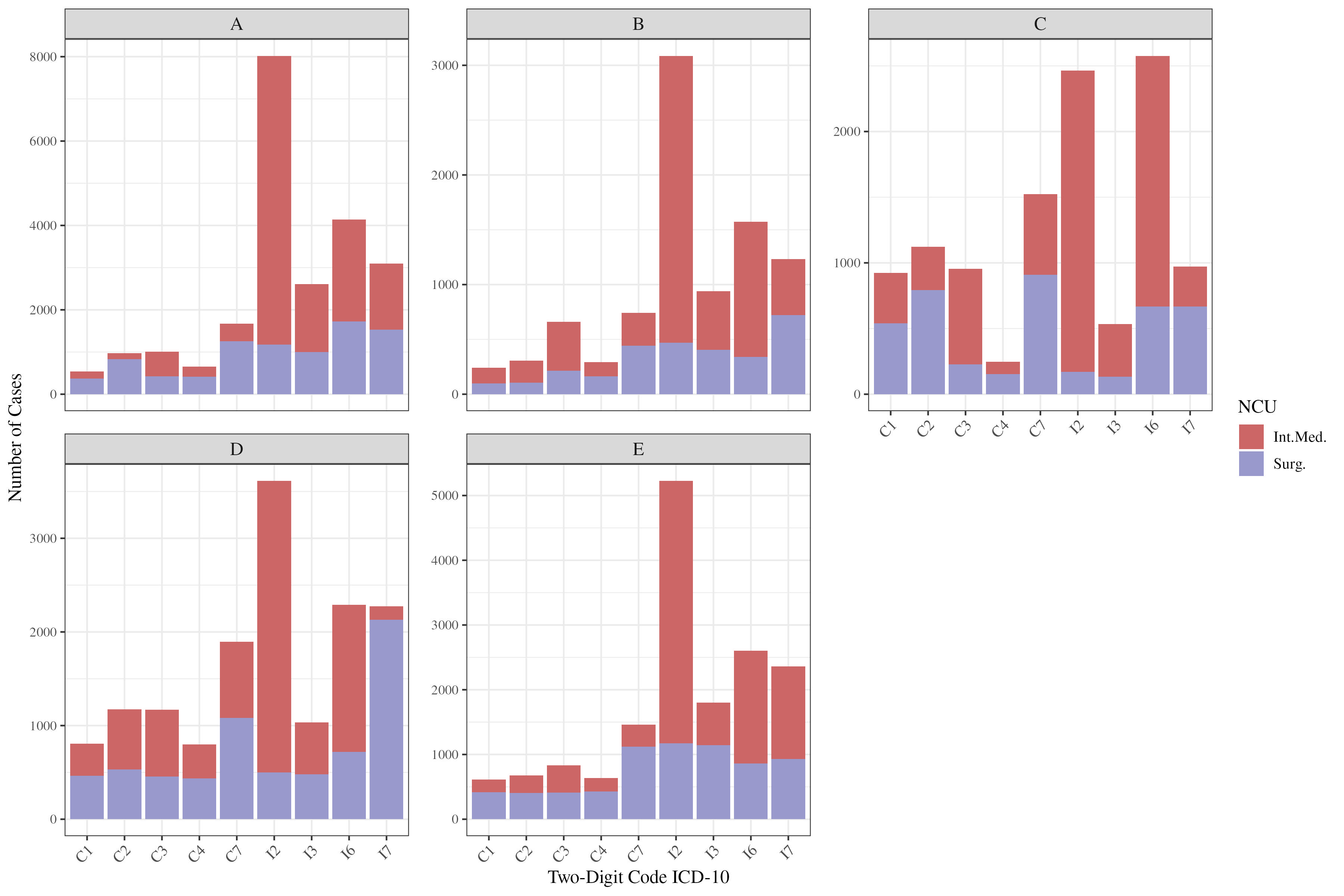}
    \caption{Distribution of Patients Across Diagnosis Groups and NCUs}
    \label{fig:panel_stacked_bar}
\end{figure}

Figure \ref{fig:panel_stacked_bar} displays the distribution of patients across ICD-10 diagnosis groups by NCU type. Cardiovascular conditions (I2, I3, I6, I7) show strong specialization toward internal medicine NCUs, while oncological diagnoses (C1-C4, C7) exhibit more balanced placement patterns. The variation in placement across diagnosis groups reflects differences in clinical requirements and treatment pathways.

\begin{figure}[htp]
    \centering
    \includegraphics[width=0.8\textwidth]{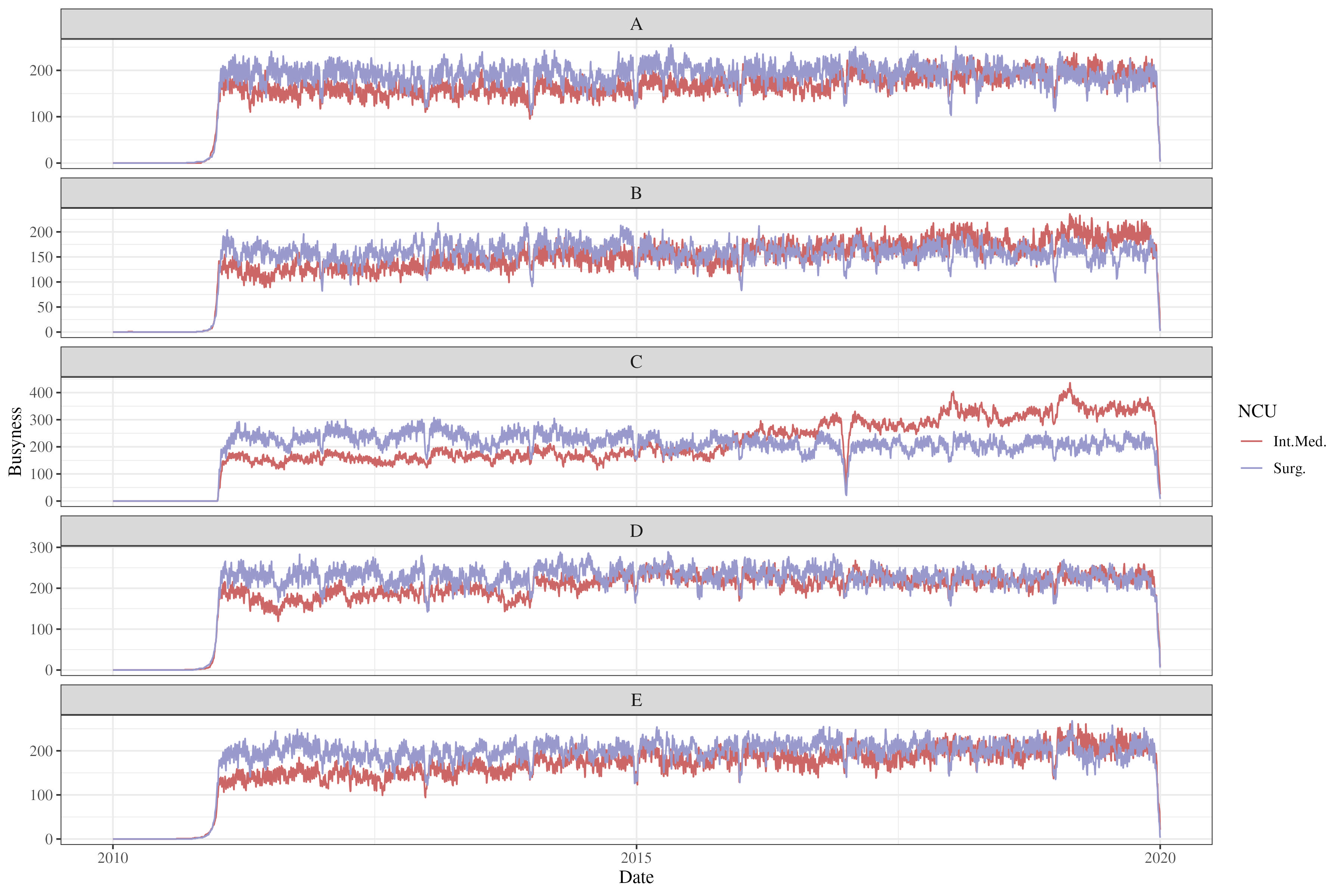}
    \caption{Busyness Time Series}
    \label{fig:busyness_timeseries}
\end{figure}

Figure \ref{fig:busyness_timeseries} shows temporal variation in NCU occupancy levels across the study period (2012-2020). Both internal medicine and surgical NCUs exhibit substantial day-to-day fluctuations in patient counts, with periodic peaks reflecting seasonal admission patterns. The observed variation in busyness levels motivates the policy framework's explicit consideration of utilization effects on patient outcomes.

\subsection{Instrument Strength}

\begin{table}[ht]
\centering
\begin{tabular}{lll}
  \hline
Model & First Stage R2 & First Stage F-statistic \\ 
  \hline
Linear Model & \LinearModelFirstStageRSQ & \LinearModelFStat \\ 
  Causal Forest & \CausalForestFirstStageRSQ &  \\ 
   \hline
\end{tabular}
\caption{Comparison of First Stage Metrics for Linear Model and Instrumental Forest }
\label{tab:IV_first_stage} 
\end{table}

Table \ref{tab:IV_first_stage} presents first-stage metrics comparing the Linear Model (LM) and Causal Forest (CF) approaches. For the Linear Model, the First Stage R-squared is \LinearModelFirstStageRSQ{} and the First Stage F-statistic is \LinearModelFStat{}. For the Causal Forest, the First Stage R-squared is \CausalForestFirstStageRSQ{}. While the instrument strength is moderate rather than exceptional, the F-statistic is larger conventional weak instrument thresholds, and controlling for patient and hospital characteristics helps ensure valid causal identification.

\subsection{Treatment Effects}

\begin{table}[ht]
\centering
\begin{tabular}{lll}
  \hline
Model & Average Treatment Effect & Standard Error \\ 
  \hline
IV Causal Forest & \ATECausalForest & \SECausalForest \\ 
  Linear IV & \ATELinearModel & \SELinearModel \\ 
   \hline
\end{tabular}
\caption{Average Treatment Effect} 
\label{tab:ate}
\end{table}

Table \ref{tab:ate} presents average treatment effect (ATE) estimates comparing the Causal Forest and Linear Model approaches. The Causal Forest estimates an ATE of \ATECausalForest{} (SE = \SECausalForest{}), while the Linear Model yields \ATELinearModel{} (SE = \SELinearModel{}). These aggregate effects mask substantial heterogeneity across patients, as demonstrated by the conditional treatment effect analysis below. The ATE reflects the average effect for the full sample but is of limited practical relevance given the extensive heterogeneity in treatment effects by patient characteristics, diagnosis, and NCU busyness levels.

\subsubsection*{Conditional Average Treatment Effect: Specialization Effect}

Table \ref{tab:cates} shows that treatment effects vary across ICD10 groups, reflecting differences in clinical specialisation between internal medicine and surgical NCUs. Cardiovascular diagnoses such as I2 and I6 show negative effects when assigned away from their specialised unit, consistent with the importance of concentrated expertise and established clinical pathways. Oncological groups display smaller and more variable effects. The pattern supports the interpretation that patient placement aligned with the appropriate NCU decreases mortality, whereas deviation from specialized environments increases mortality. The standard errors are large.

\begin{table}[ht]
\centering
\begin{tabular}{rrr}
  \hline
 & CATE & SE \\ 
  \hline
I2 & -0.16 & 0.09 \\ 
  I3 & 0.12 & 0.14 \\ 
  I6 & -0.64 & 0.26 \\ 
  I7 & -0.28 & 0.14 \\ 
  C1 & -0.02 & 0.36 \\ 
  C2 & 0.13 & 0.32 \\ 
  C3 & -0.30 & 0.32 \\ 
  C4 & -0.09 & 0.24 \\ 
  C7 & 0.14 & 0.19 \\ 
   \hline
\end{tabular}
\caption{CATE estimates with standard errors} 
\label{tab:cates}
\end{table}

\subsubsection*{Conditional Average Treatment Effects: Utilization Effect}
\begin{figure}[htp]
    \centering
    \includegraphics[width=0.8\textwidth]{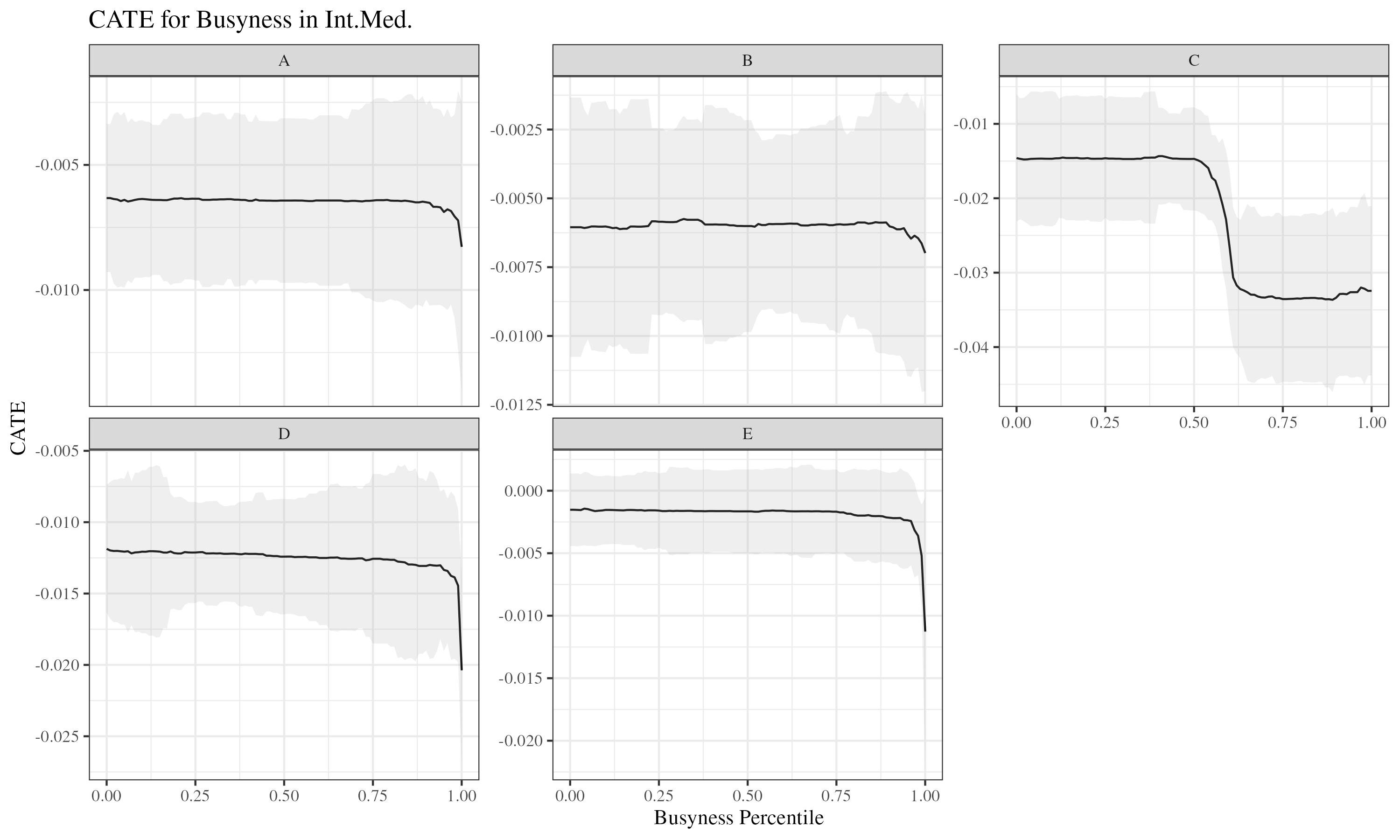}
    \caption{Conditional Average Treatment Effects by NCU Busyness and Hospital}
    \label{fig:cate_busyness_main}
\end{figure}

Figure \ref{fig:cate_busyness_main} shows that treatment effects vary with occupancy levels. For both internal medicine and surgical NCUs, rising busyness is associated with declining outcomes. The gradient is particularly pronounced in hospitals with recurrent peak occupancy, indicating that capacity pressure interferes with timely interventions and care continuity. 

The variation across hospitals suggests that utilisation effects depend on institutional characteristics such as staffing resilience, workflow organisation, and specialist availability. These utilisation-related declines in treatment effects show that operational conditions interact with clinical decision making. This reinforces the relevance of modelling placement as a function of real-time capacity rather than relying solely on clinical classification.

The evidence confirms that evaluating policies without reference to busyness levels would overlook a central driver of outcome variation.

Figure \ref{fig:cate_busyness_main} displays conditional average treatment effects as a function of NCU busyness levels, grouped by hospital.Across most hospitals, higher busyness levels are associated with declining treatment benefits, consistent with capacity constraints reducing care quality. However, substantial heterogeneity emerges across hospitals and busyness levels. For internal medicine NCUs, the negative relationship between busyness and outcomes is most pronounced in Hospitals C and E, where extremely high occupancy substantially reduces survival probabilities. Surgical NCUs exhibit similar patterns but with greater variation across hospitals. These findings demonstrate that treatment effects are conditional on operational circumstances, validating the policy framework's emphasis on balancing specialization benefits against utilization costs. The heterogeneity across hospitals underscores the importance of tailoring placement policies to local contexts.

\subsection{Bounds}

\begin{table}[ht]
\centering
\begin{tabular}{rrr}
  \hline
 & Lower Bound & Upper Bound \\ 
  \hline
Frequentist & -0.13 & 0.00 \\ 
  Manski & -0.51 & 0.49 \\ 
  Pearl & -0.05 & 0.03 \\ 
   \hline
\end{tabular}
\caption{ATE Bound Estimates} 
\label{tab:ate_bounds}
\end{table}

Table \ref{tab:ate_bounds} presents the estimated bounds for the average treatment effect. The frequentist bounds are relatively narrow and and vary between patients and effect sizes. Manski bounds allow for weaker assumptions and therefore produce a wider interval. Pearl bounds, which incorporate the structure of the instrument, tighten the region of feasible effects and reduce uncertainty relative to the Manski set. All three appraoche sto bounds contain the zero effect. THis motivates us to use minimax regret framework, which selects policies that remain defensible across the entire admissible range of treatment effects.

The width and patient-level variation of these bounds determines their operational utility. Manski bounds, while theoretically conservative, converge to nearly uniform intervals across patients (approaching [0,1] for most observations), offering minimal discriminatory power for policy differentiation. In contrast, Pearl bounds and Frequentist confidence sets maintain sufficient heterogeneity to distinguish optimal assignments while remaining robust to misspecification. This patient-level variation is essential for operational policy implementation, as uniform bounds provide no guidance for differential treatment assignment.

\begin{figure}[ht]
    \centering
    \begin{subfigure}{0.28\linewidth}
        \centering
        \includegraphics[width=\linewidth]{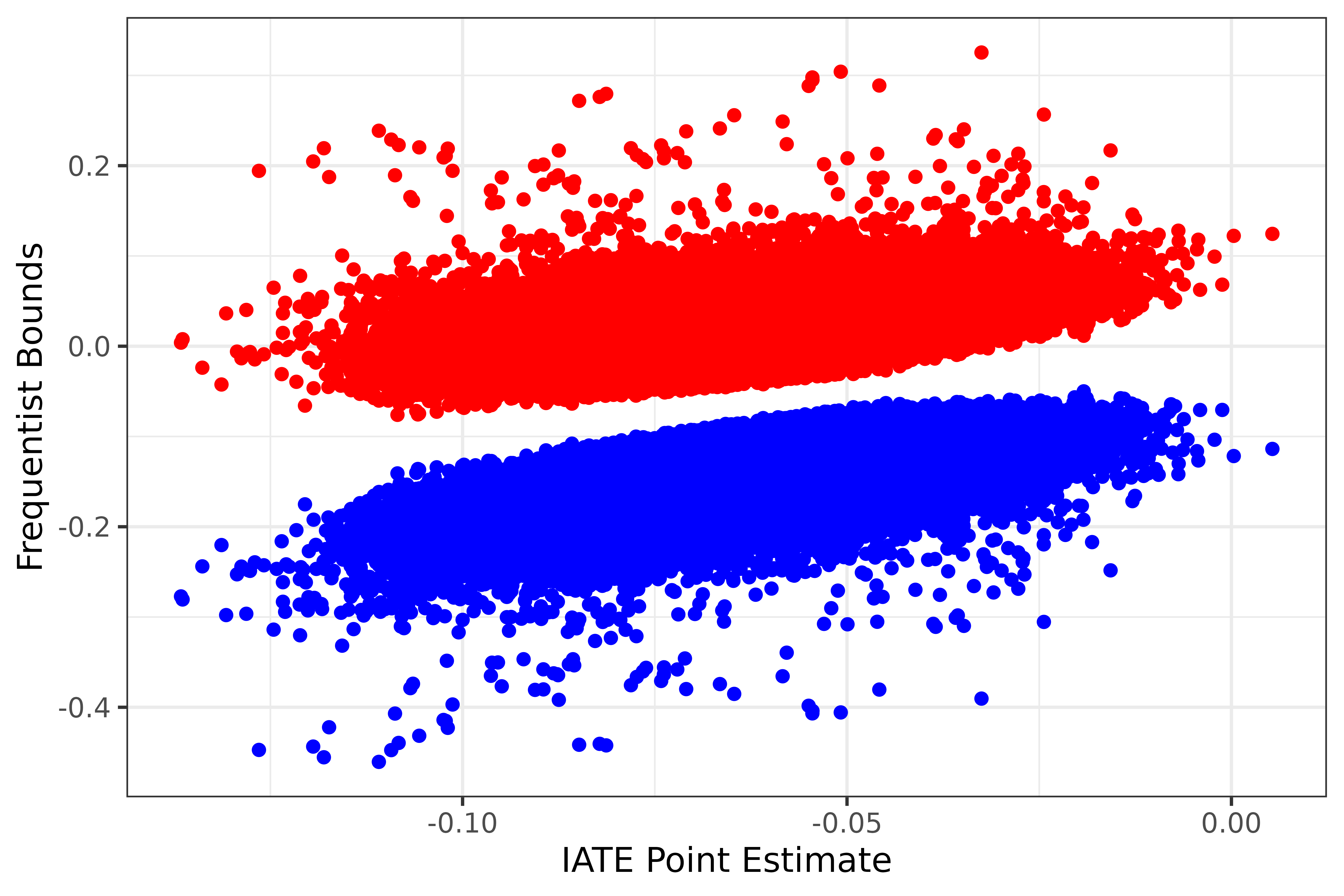}
        \caption{Frequentist Bounds}
        \label{fig:freq_bounds}
    \end{subfigure}
    \hfill
    \begin{subfigure}{0.28\linewidth}
        \centering
        \includegraphics[width=\linewidth]{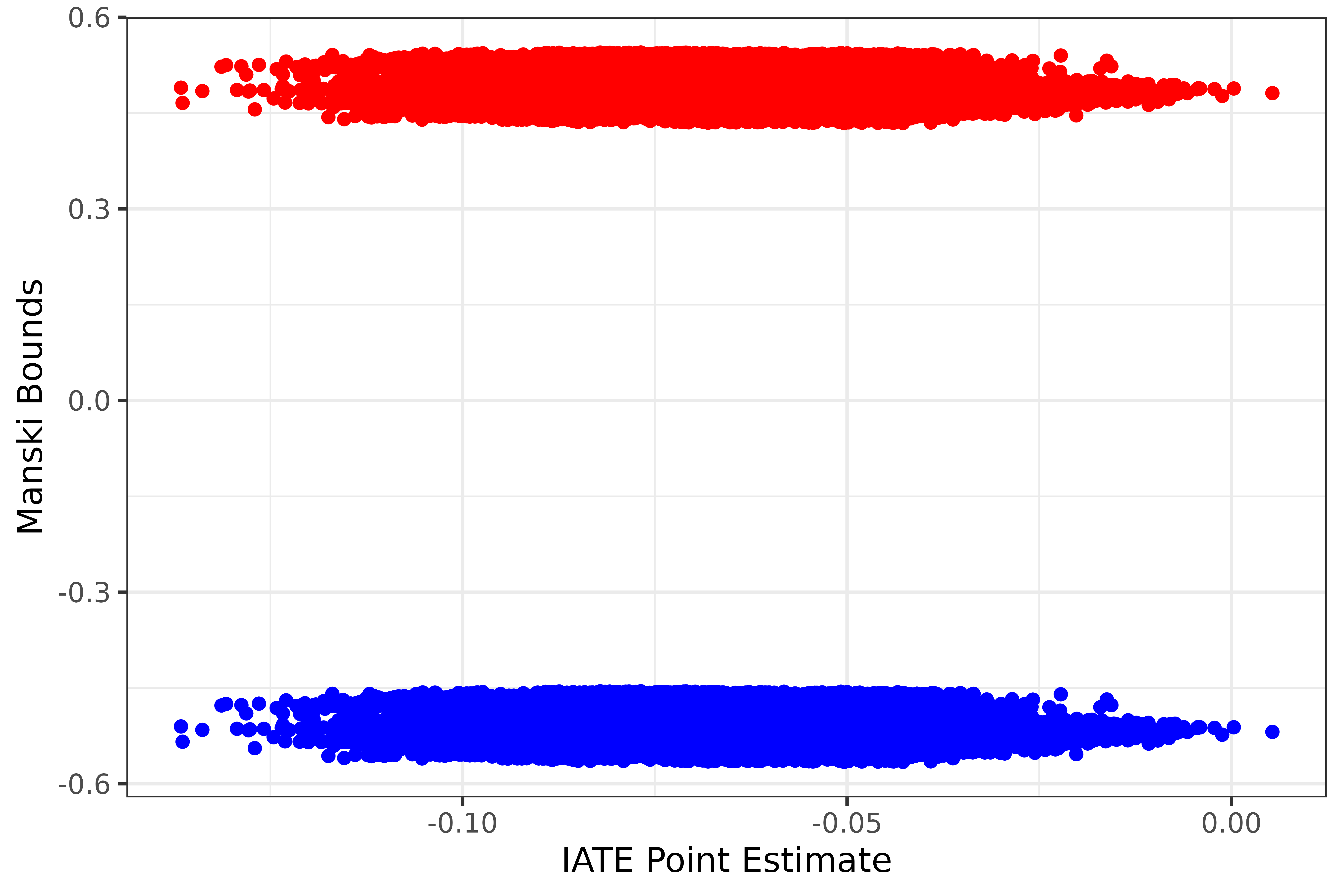}
        \caption{Manski Bounds}
        \label{fig:manksi_bounds}
    \end{subfigure}
    \hfill
    \begin{subfigure}{0.28\linewidth}
        \centering
        \includegraphics[width=\linewidth]{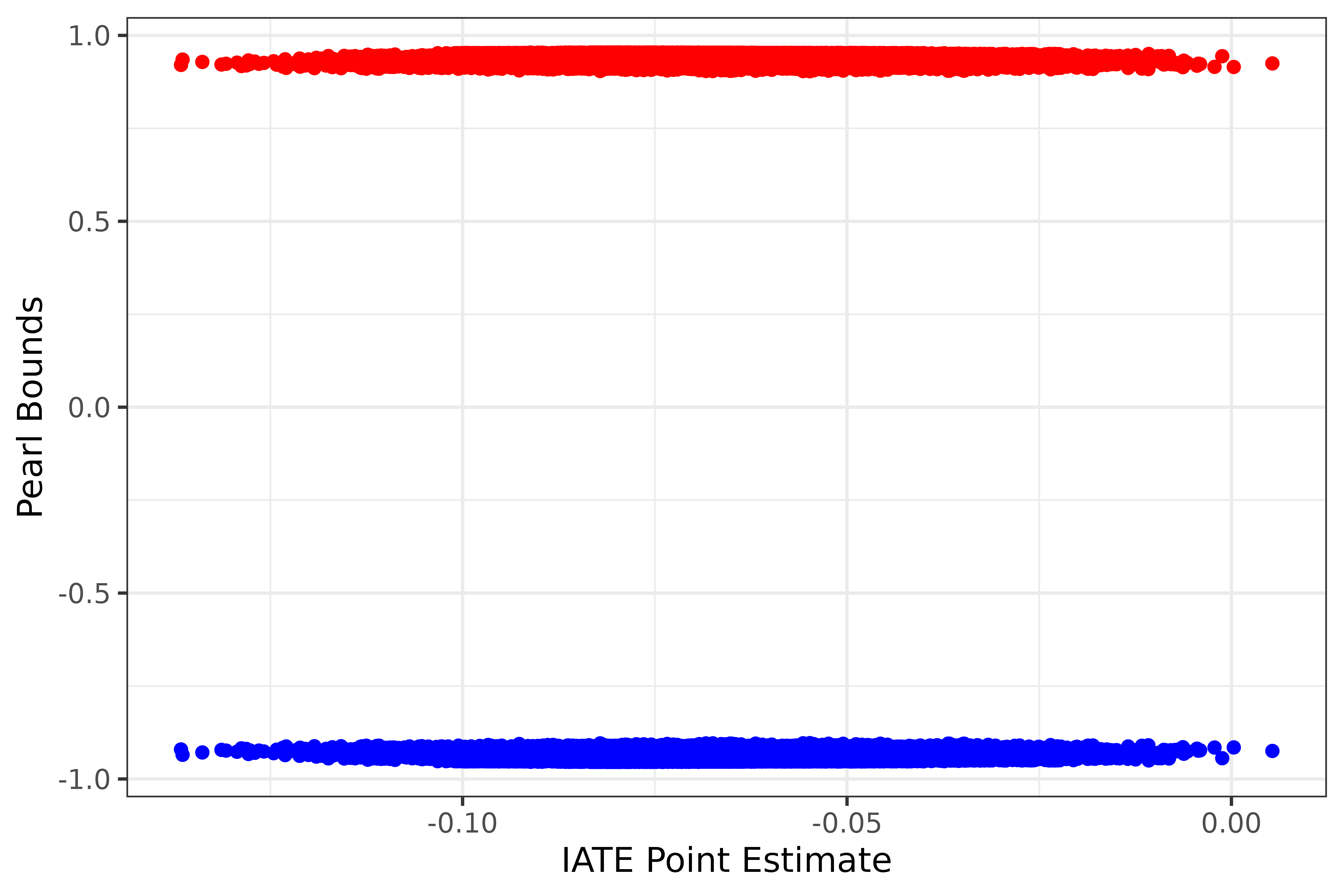}
        \caption{Pearl Bounds}
        \label{fig:pearl_bounds}
    \end{subfigure}
    
    \caption{IATE Bounds Comparision}
    \label{fig:bounds_comparison}
\end{figure}

Figure \ref{fig:bounds_comparison} illustrates how the width of the admissible effect region differs across the three approaches. The frequentist bounds remain concentrated around the point estimate, Manski bounds expand the feasible outcome space due to weaker assumptions, and Pearl bounds reduce this space by incorporating the structure of the instrument. The visual comparison highlights how methodological choices translate into different levels of uncertainty and therefore shape the range of defensible policy decisions.

\subsection{Policy Evaluation Results}

\begin{table}[ht]
\centering
\caption{Policy Performance Without Reassignment Constraints} 
\label{tab:unconstrained_welfare}
\begin{tabular}{rrrrr}
  \hline
 & Welfare Max & Frequentist Minimax & Manski Minimax & Pearl Minimax \\ 
  \hline
Welfare & 0.00373 & 0.00363 & 0.00373 & 0.00359 \\ 
  Frequentist Regret & 0.00036 & 0.00015 & 0.00036 & 0.00049 \\ 
  Manski Regret & 0.00000 & 0.00010 & 0.00000 & 0.00014 \\ 
  Pearl Regret & 0.00049 & 0.00051 & 0.00049 & 0.00034 \\ 
   \hline
 \hline
\end{tabular}
\end{table}


Table \ref{tab:unconstrained_welfare} presents policy evaluation results for the unconstrained scenario. Rows display evaluation criteria (welfare gains in row 1, regret measures in rows 2 through 4), while columns show different decision rules (welfare maximization, frequentist minimax regret, Manski minimax regret, and Pearl minimax regret). Diagonal elements represent optimal performance when the decision rule matches the evaluation metric. Off-diagonal elements show how each policy performs when evaluated by alternative criteria. Results for scenarios with 10\% and 20\% reassignment limits are shown in Tables \ref{tab:10pct_welfare} and \ref{tab:20pct_welfare} in the appendix.

We find that welfare maximization achieves an expected gain of 0.00373 additional survivors per patient under optimal assignment, corresponding to approximately 3.7 additional survivors per 1,000 patient placements. Given the baseline survival rate of 97\% (in-hospital mortality rate of 3\%), this is a relative reduction in mortality risk of approximately 12.3\% compared to observed placements. These gains arise from optimized reallocation of patients across existing NCU capacity without requiring infrastructure investment, capacity expansion, or additional staffing resources.

The Manski minimax regret policy exhibits zero regret when evaluated against its own metric and produces identical welfare to the welfare maximization policy. This convergence happens because the individualized bounds lack sufficient variation across patients. When lower bounds provide no discriminatory power to differentiate between patients, the minimax regret problem reduces to welfare maximization. We find that bounds must vary meaningfully across patients to function effectively in this decision context, rather than simply being tight in absolute terms. Manski bounds, while theoretically conservative, converge to near-uniform intervals across the patient population, eliminating their utility for patient-level discrimination. Pearl bounds and frequentist confidence intervals preserve heterogeneity across patients, enabling differentiation between optimal assignments.

The frequentist minimax regret policy demonstrates the tradeoff between robustness and welfare maximization. When evaluated by frequentist regret, the welfare maximization policy achieves a regret of 0.00036, while the frequentist minimax regret policy reduces this to 0.00015. This reduction of 0.00021 (from 0.00036 to 0.00015) represents improved protection against worst-case scenarios: the frequentist minimax regret policy reduces the maximum opportunity cost by 0.00021 survivors per patient when treatment effects fall at the lower bounds of the confidence intervals rather than at point estimates. In practical terms, this means that under pessimistic realizations of treatment effects (lower confidence bound scenario), the frequentist minimax regret policy prevents approximately 0.3 additional deaths per 1,000 patients compared to welfare maximization. This robustness comes at a cost, because frequentist minimax regret policy achieves welfare of 0.00363 survivors per patient compared to 0.00373 for welfare maximization, sacrificing 0.00010 survivors per patient (approximately 0.1 fewer survivors per 1,000 patients, or roughly 3\% of the potential welfare gain). The small magnitude of this tradeoff indicates that hedging against uncertainty through minimax regret imposes minimal welfare losses while providing meaningful protection against worst-case scenarios using the frequentist bounds.

\subsubsection*{Policy Effects on Utilisation and Specialisation}

The welfare maximization policy achieves mortality reductions through two operational mechanisms: balancing NCU utilization across units and aligning with clinical specialization patterns. We examine how the policy affects both dimensions relative to observed hospital placements. Tables \ref{tab:utilisation_intmed} and \ref{tab:utilisation_surgical} presents NCU busyness levels under observed and optimized policy placements. Results show that the policy achieves welfare improvements through modest, balanced reallocation rather than substantial shifts in utilization patterns. Across all hospitals, mean busyness changes minimally, with more patients placed in surgical NCUs. This pattern reflects the relatively large negative ATE and CATE shown in Table \ref{tab:ate} and Table \ref{tab:cates}.

\begin{table}[ht]
\centering
\caption{Internal Medicine NCU Utilisation: Observed vs Welfare Max Policy} 
\label{tab:utilisation_intmed}
\begin{tabular}{rrrrr}
  \hline
 & Obs Mean & Policy Mean & Obs Max & Policy Max \\ 
  \hline
A & 183.43 & 182.99 & 238 & 238 \\ 
  B & 176.32 & 175.84 & 231 & 231 \\ 
  C & 243.08 & 242.64 & 417 & 417 \\ 
  D & 215.64 & 215.25 & 265 & 265 \\ 
  E & 193.06 & 192.60 & 253 & 253 \\ 
   \hline
\end{tabular}
\end{table}


\begin{table}[ht]
\centering
\caption{Surgical NCU Utilisation: Observed vs Welfare Max Policy} 
\label{tab:utilisation_surgical}
\begin{tabular}{rrrrr}
  \hline
 & Obs Mean & Policy Mean & Obs Max & Policy Max \\ 
  \hline
A & 203.38 & 203.81 & 247 & 247 \\ 
  B & 174.14 & 174.61 & 218 & 218 \\ 
  C & 232.69 & 233.14 & 302 & 302 \\ 
  D & 234.93 & 235.33 & 274 & 274 \\ 
  E & 209.86 & 210.32 & 262 & 262 \\ 
   \hline
\end{tabular}
\end{table}


Hospital-level analysis shows shifts in mean busyness with more patients placed in surgical NCUs. Hospital A shows minor shifts (internal medicine: from 183.43to 182.99; surgical: from 203.38 to 203.82), while Hospital C shows similar marginal adjustments (internal medicine: from 243.08 to 242.64; surgical: from 232.69 to 233.14). Maximum busyness levels remain unchanged as defined by capacity constraints. The policy requires no additional hospital capacity, as welfare gains are achieved through optimized assignment of existing resources. The balanced redistribution avoid new capacity bottlenecks while using underutilized capacity in less congested units. Table \ref{tab:specialisation_by_diagnosis} examines placement frequency in clinically specialized NCUs under observed hospital assignments versus the four policy variants.

\begin{table}[ht]
\centering
\caption{Share of patients placed into the Int. Med. NCU} 
\label{tab:specialisation_by_diagnosis}
\begin{tabular}{rrrrrr}
  \hline
 & Observed & Welfare Max & Frequentist & Manski & Pearl \\ 
  \hline
C1 & 0.40552 & 0.43448 & 0.50207 & 0.43448 & 0.58897 \\ 
  C2 & 0.37132 & 0.41216 & 0.48528 & 0.41216 & 0.55176 \\ 
  C3 & 0.64716 & 0.45783 & 0.50000 & 0.45783 & 0.55422 \\ 
  C4 & 0.39024 & 0.48943 & 0.42114 & 0.48943 & 0.53496 \\ 
  C7 & 0.33797 & 0.56420 & 0.57977 & 0.56420 & 0.58755 \\ 
  I2 & 0.81476 & 0.45766 & 0.48468 & 0.45766 & 0.51169 \\ 
  I3 & 0.52660 & 0.43933 & 0.48117 & 0.43933 & 0.47280 \\ 
  I6 & 0.63684 & 0.34749 & 0.40692 & 0.34749 & 0.42521 \\ 
  I7 & 0.45373 & 0.38669 & 0.43957 & 0.38669 & 0.49575 \\ 
   \hline
\end{tabular}
\end{table}


Under observed placements, hospitals show strong specialization patterns for cardiovascular patients: 81.5\% of acute myocardial infarction patients (I2) are placed in internal medicine NCUs, as are 63.7\% of cerebrovascular disease patients (I6). Specialization adherence is lower for chronic ischemic heart disease (I3: 52.7\%) and other circulatory diseases (I7: 45.4\%). Oncology patients show varied placement patterns across NCU types, ranging from 33.8\% to 64.7\% in internal medicine.

Policy variants show different placement patterns compared to observed assignments. Welfare maximization places 45.8\% of I2 patients in internal medicine, compared to 81.5\% under observed assignments. Similar changes occur for I3 (43.9\% vs 52.7\%), I6 (34.7\% vs 63.7\%), and I7 (38.7\% vs 45.4\%). The Frequentist, Manski, and Pearl minimax regret policies exhibit varying specialization rates across diagnoses, with Pearl showing higher internal medicine placement rates ranging from 42.5\% to 58.9\% across diagnoses. Oncology patients show varied placement patterns across policies.

The difference between observed and policy-driven placements indicates that the estimated treatment effects favor surgical NCU placement for most patients in the sample. Three factors may explain this pattern. First, the instrumental variable identification strategy isolates variation in placement driven by emergency admission shocks, which may disproportionately affect internal medicine NCU congestion. If internal medicine units experience larger busyness penalties during high-volume periods, estimated treatment effects favor surgical placement to avoid utilization-related mortality increases. Second, the sample restriction to ICD-10 codes I2-I7 and C1-C4, C7 may select patients whose clinical profiles are appropriate for both NCU types, reducing the marginal benefit of specialized placement. Third, if observed placements overweight historical specialization norms relative to actual outcome differences, the data-driven policy may correct for inefficient adherence to traditional assignment rules. 

Table \ref{tab:welfare_by_diagnosis} quantifies the contribution of each diagnosis to aggregate welfare gains. Cardiovascular diagnoses generate the largest absolute welfare gains: I2 patients contribute 13.6 additional survivors (average gain: 0.00309 per patient), followed by I6 (3.8 total, 0.00248 per patient), I7 (3.6 total, 0.00170 per patient), and I3 (3.3 total, 0.00197 per patient). Oncology diagnoses contribute smaller gains: C3 (2.8 total, 0.00242 per patient), C7 (2.0 total, 0.00109 per patient), C2 (1.4 total, 0.00133 per patient), C1 (1.1 total, 0.00146 per patient), and C4 (0.8 total, 0.00133 per patient). 

These per-patient welfare gains, ranging from 0.00109 to 0.00309, represent mortality reductions achieved without capital investment or capacity expansion, which is relevant for resource-constrained healthcare systems.

\begin{table}[ht]
\centering
\caption{Welfare Gains by Diagnosis Group (Welfare Max.)} 
\label{tab:welfare_by_diagnosis}
\begin{tabular}{rrrr}
  \hline
 & N & Avg Welfare Gain & Total Welfare Gain \\
  \hline
I2 &   4405 & 0.00309 & 13.61960 \\
  I6 &   1531 & 0.00248 & 3.79047 \\
  I7 &   2118 & 0.00170 & 3.60302 \\
  I3 &   1673 & 0.00197 & 3.29019 \\
  C3 &   1162 & 0.00242 & 2.81747 \\
  C7 &   1799 & 0.00109 & 1.96172 \\
  C2 &   1053 & 0.00133 & 1.39870 \\
  C1 &    725 & 0.00146 & 1.06166 \\
  C4 &    615 & 0.00133 & 0.81580 \\
   \hline
\end{tabular}
\end{table}


Reassignment rates show that cardiovascular patients are more likely to be moved than oncology patients: I2 (77.3\%), I6 (63.9\%), I7 (45.6\%), and I3 (53.0\%) compared to C3 (62.8\%), C7 (35.9\%), C2 (39.2\%), C1 (42.9\%), and C4 (42.1\%). The high reassignment rates for I2 and I6 align with the specialization analysis, indicating that the policy favors surgical NCU placement for many cardiovascular patients when internal medicine units are congested. Per-patient welfare gains are relatively uniform across diagnoses, ranging from 0.00109 to 0.00309, suggesting that heterogeneity in total contributions primarily reflects sample size differences rather than varying treatment effect magnitudes. Across the 15,081 patients in the test period, the welfare maximization policy would prevent approximately 69 deaths compared to observed hospital placements. The policy achieves these welfare improvements through balanced utilization adjustments without violating capacity constraints.

\section{Implications}

The empirical findings generate several actionable implications for hospital operations, clinical practice, and health policy. The welfare maximization policy achieves 0.00373 additional survivors per patient compared to observed placements, corresponding to approximately 69 prevented deaths across 15,081 test patients. While this absolute magnitude appears modest, these gains arise purely from optimized reallocation of existing NCU capacity, requiring no infrastructure investment, capacity expansion, or additional staffing. For resource-constrained healthcare systems, mortality reductions achieved through improved operational efficiency represent a practical pathway to improved outcomes without capital expenditure.

However, implementation must address two critical tensions revealed by the analysis. First, the policy substantially reduces adherence to traditional specialization patterns: acute myocardial infarction patients (I2) placed in internal medicine NCUs decline from 81.5\% under observed assignments to 45.8\% under the optimal policy, with similar reductions across cardiovascular diagnoses. This dramatic departure from clinical norms suggests either that utilization costs dominate specialization benefits when NCUs are congested, or that unobserved confounding biases the estimated treatment effects. Second, the computational complexity of jointly optimizing patient assignments under capacity constraints limits scalability: exact optimization becomes infeasible beyond 22 patients per day, requiring greedy approximations that do not guarantee global optimality. These methodological and operational constraints shape implementation strategies.

\subsection*{Implications for Patients}
For patients, data-driven placement policies offer mortality reductions through two mechanisms: avoiding congested NCUs where high occupancy degrades care quality, and ensuring timely access to appropriate clinical expertise. The policy achieves a 12.3\% relative reduction in mortality risk compared to observed placements. Patient placement decisions have documented effects on patient outcomes and satisfaction, with multiple factors influencing patient experience in hospital ward settings \citep{guan2024factors}. However, the substantial deviation from traditional specialization patterns raises important questions about clinical safety and patient preferences. Cardiovascular patients accustomed to internal medicine NCU care may experience concern or confusion if assigned to surgical NCUs, even when data suggest equivalent or superior outcomes under those operational conditions. Transparent communication about placement rationale, continuous outcome monitoring, and provisions for clinical override in high-risk cases will be essential to maintain patient trust and safety during implementation.

\subsection*{Implications for Providers}
For hospital administrators and clinical staff, the findings offer both opportunities and operational challenges. The policy demonstrates that meaningful welfare gains are achievable through optimized bed management without capacity expansion, directly addressing resource allocation pressures common in acute care settings. The estimated treatment effects reveal that utilization costs (NCU congestion effects) rival or exceed specialization benefits for many patient types, suggesting that real-time busyness monitoring should inform placement decisions rather than relying solely on diagnosis-specialty matching. Clinical decision support systems have demonstrated success in improving clinical practice when properly designed and implemented \citep{kawamoto2005improving}, with information technology initiatives showing significant potential for enhancing patient safety \citep{bates2003improving}.

However, prudent implementation requires staged validation rather than immediate full deployment. We recommend a three-phase approach: (1) apply the policy initially to patients with low baseline specialization adherence (e.g., chronic ischemic heart disease I3, other circulatory diseases I7), where traditional clinical norms provide weak guidance and reassignment poses minimal disruption to established care pathways; (2) monitor outcomes closely for 6-12 months, comparing realized mortality rates against predicted welfare gains to validate that estimated treatment effects translate to real-world improvements; (3) expand to patients with stronger traditional specialization patterns (e.g., acute myocardial infarction I2, cerebrovascular disease I6) only if empirical validation confirms welfare improvements without adverse clinical consequences. This conservative approach balances the potential for mortality reduction against risks of model misspecification or unobserved confounding.

The substantial hospital-specific heterogeneity observed in treatment effects and utilization patterns underscores the need for local calibration rather than uniform policy deployment. Hospital C, with the highest baseline occupancy, exhibits pronounced negative relationships between busyness and outcomes, suggesting that congestion management should dominate placement decisions in high-volume settings. Hospitals with lower baseline occupancy may prioritize specialization alignment more heavily. Providers should invest in hospital-specific analytics capabilities to estimate local treatment effects, monitor real-time NCU busyness, and enable adaptive decision support systems that update placement recommendations based on observed outcomes.

The computational complexity of exact optimization (infeasible beyond 22 patients per day, affecting 68.4\% of test days) requires greedy algorithms that provide tractable approximations but sacrifice guaranteed optimality. Future work should investigate computational improvements, including mixed-integer programming formulations, parallel evaluation architectures, or machine learning approaches to policy approximation. Reassignment constraints further limit achievable welfare: restricting changes to 10\% or 20\% of patients recovers only 26\% and 36\% of unconstrained welfare gains, highlighting the importance of operational flexibility. Hospitals should evaluate feasible reassignment rates within their organizational contexts and calibrate policies accordingly.

\subsection*{Methodological Implications}
The analysis shows a requirement for partial identification approaches in treatment assignment problems: bounds must exhibit sufficient patient-level variation to enable discriminatory treatment assignment, not merely be tight in absolute terms. Manski worst-case bounds, while theoretically conservative and making minimal assumptions, converge to nearly uniform intervals approaching [0,1] across patients, eliminating their utility for policy differentiation. When bounds lack discriminatory power, minimax regret policies collapse to welfare maximization regardless of theoretical conservatism. Pearl bounds and frequentist confidence intervals preserve some heterogeneity across patients, enabling meaningful policy differentiation even when they ultimately align closely with welfare maximization in assignment decisions.

This finding has implications beyond hospital operations. In any sequential or joint decision problem where partial identification methods inform treatment assignment, the operational value of a bounding approach depends critically on its ability to distinguish between decision units (patients, customers, participants), not solely on interval width or coverage guarantees. Future methodological work on individualized bounds should prioritize maintaining patient-level heterogeneity alongside theoretical conservatism, as uniform bounds provide no actionable guidance for differential treatment assignment regardless of their width or validity.

\section{Limitations}

Several limitations should be acknowledged:

(1) the dataset lacks direct information on staff availability and real-time resource constraints within NCUs. The true level of unit utilization and its impact on patient outcomes may not be fully captured, and unobserved operational factors could influence the results. Additionally, we restrict the analysis to five university hospitals and selected diagnosis groups (I2, I3, I6, I7, C1 to C4, C7), which may limit generalizability to other hospitals or patient populations.

(2) while we control for NCU busyness levels to address potential SUTVA violations, these controls may not fully capture spillover effects between patients. The busyness measure provides partial control for congestion effects, but unmeasured aspects of care quality under varying workload conditions could still lead to interference between patient outcomes. We interpret the estimated treatment effects as conditional on observed NCU utilization levels \citep{rubin1974estimating, athey2019machine}.

(3) the instrumental variable assumes similar effects on treatment probability across internal medicine and surgical NCUs. Differences in clinical practice, organizational structure, or local protocols could lead to deviations from this assumption. The instrument strength is moderate rather than exceptional, and causal identification depends on the validity of these assumptions.

(4) policy optimization requires evaluating feasible patient assignments under capacity constraints, creating computational challenges as patient counts increase. The greedy algorithm provides a tractable approximation but does not guarantee global optimality. In future, research I could analyse how big the difference is between greedy and exact estimation.

(5) the frequentist confidence intervals we use are not bounds in the formal sense. They quantify sampling uncertainty around point estimates rather than identification regions under structural assumptions. We include them in the minimax regret framework because they provide patient-level discrimination necessary for differentiated treatment assignment, but they capture estimation error rather than structural ambiguity.

(6) we apply Manski and Pearl bounds at the patient level by conditioning on observed covariates, though these methods were originally developed for population-level partial identification. More appropriate approaches for constructing individualized bounds that better account for patient-specific uncertainty may exist, and methodological extensions could improve the precision and interpretability of patient-level bounds in future work.

(7) the policy recommendations show substantial deviations from observed specialization patterns, which raises questions about clinical appropriateness. Two interpretations should be considered. First, if the estimated treatment effects accurately reflect causal impacts, the results suggest that utilization costs dominate specialization benefits for most patients in congested settings, implying that hospitals should prioritize busyness management over traditional specialization rules. Second, if unobserved confounding or model misspecification biases the estimates, the policy recommendations may underweight the true benefits of specialized care, this would require a very careful implementation. Conservative implementation would involve step-wise validation: initially applying the policy selectively to patients with borderline placement decisions (e.g., I3 and I7 with low baseline specialization), monitoring outcomes closely, and gradually expanding to patients with stronger traditional specialization patterns (e.g., I2 and I6) only if empirical validation confirms welfare improvements without adverse clinical consequences.

Despite these limitations, the approach provides a meaningful step toward data-driven patient placement policies and offers practical insights for optimizing clinical outcomes and resource utilization in hospital settings.

\section{Conclusion}
We integrate instrumental variable causal forests with a minimax regret policy framework to optimize patient placement in normal care units, using administrative data from Swiss university hospitals. The analysis reveals two central mechanisms: a utilization effect, where higher NCU occupancy worsens outcomes, and a specialization effect, where clinical expertise match improves survival. By leveraging exogenous variation in daily emergency admissions as an instrument, we estimate heterogeneous treatment effects that balance these competing forces.

The policy achieves mortality reductions through optimized reallocation without requiring capacity expansion or infrastructure investment. We show that welfare maximization, and frequentist confidence intervals, to some extend also Pearl bounds, produce destinctive assignments because these approaches maintain sufficient patient-level discriminatory power to distinguish optimal placements. In contrast, Manski bounds converge to nearly uniform intervals across patients, eliminating their ability to differentiate assignments despite being theoretically conservative. This finding shows that mini-max regret policies using bounds require patient-level variation to enable discriminatory treatment assignment, not just tight bounds.

Computational constraints from jointly optimizing patient assignments under capacity limits necessitate greedy approximations for large cohorts. Hospital-specific heterogeneity in treatment effects and utilization patterns underscores the need for tailored implementation strategies rather than uniform policy deployment. The framework demonstrates that systematic, data-driven approaches to hospital operations can achieve mortality reductions through optimized resource allocation, offering a pathway toward more efficient healthcare delivery in resource-constrained settings.

\newpage
\addcontentsline{toc}{section}{References}
\bibliography{Bibliography.bib}
\newpage

\begin{appendix}
\setcounter{page}{1}
\pagenumbering{arabic}
\renewcommand*{\thepage}{A\arabic{page}}

\section{Appendix}\label{Appendix}

\subsection{Common Support Analysis}

\begin{figure}[htp]
    \centering
    \includegraphics[width=0.8\textwidth]{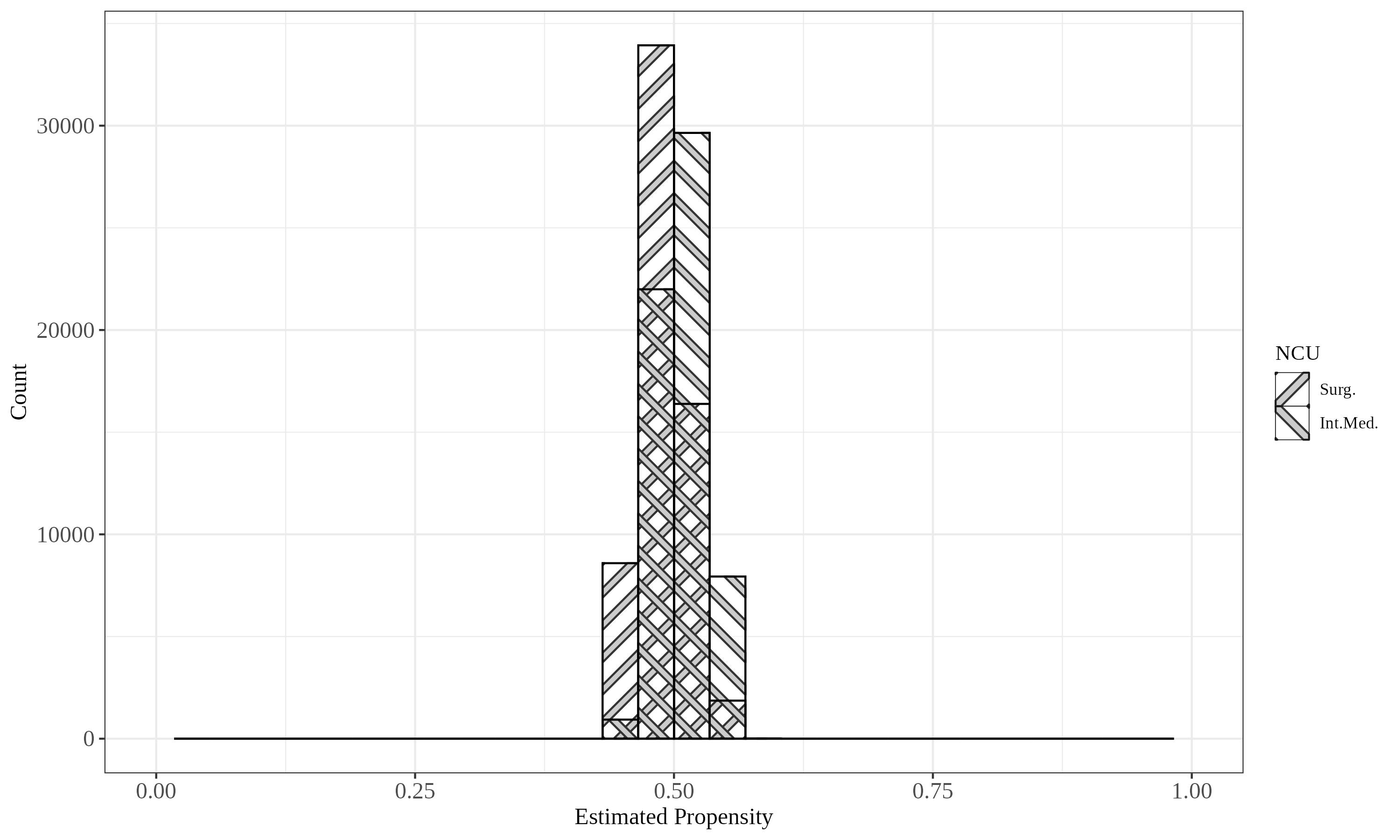}
    \caption{Common Support Plot}
    \label{fig:cs_plot}
\end{figure}

The common support plot (Figure \ref{fig:cs_plot}) shows the distribution of estimated propensity scores for patients assigned to different NCUs. The x-axis represents the estimated propensity scores, ranging from 0 to 1, and the y-axis indicates the count of patients. The plot is divided into two groups based on NCU assignment, with one group corresponding to patients assigned to internal medicine NCUs (NCU = 1) and the other group to surgical NCUs (NCU = 0). The histogram uses overlapping patterns to display the distribution of propensity scores for both groups, providing a visual representation of the common support region where the propensity scores overlap. The use of the emergency admission count as an instrument ensures that there is no strong selection between the NCUs, so common support is fulfilled.

\subsection{Compliance and Instrument Validity}

\begin{figure}[htp]
    \centering
    \includegraphics[width=0.8\textwidth]{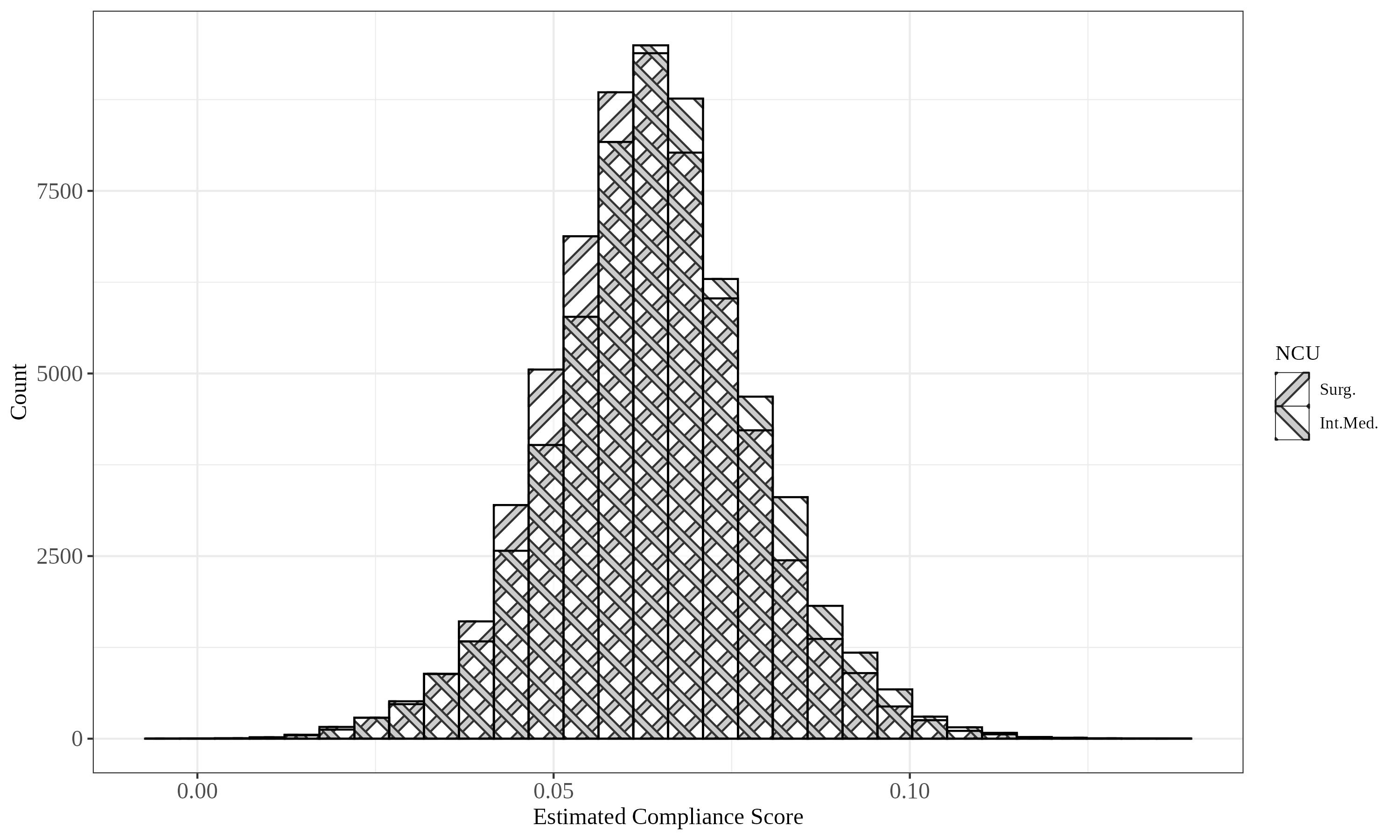}
    \caption{Compliance Score Distribution}
    \label{fig:compliance_score}
\end{figure}

The compliance score distribution (Figure \ref{fig:compliance_score}) illustrates heterogeneity in how the instrument affects treatment assignment across patients. Most observations exhibit positive compliance scores, indicating that higher emergency volumes increase the probability of internal medicine NCU assignment. A small proportion of negative values appears, likely reflecting statistical uncertainty or hospital-specific protocols where high emergency volume triggers overflow to surgical NCUs. The predominance of positive compliance supports the monotonicity assumption underlying the LATE framework.

\subsection{Reassignment Limit}

When capacity constraints limit reassignments to 10\% or 20\% of patients, welfare gains decline substantially, recovering only 26\% and 36\% of unconstrained potential welfare respectively. The convergence pattern between welfare maximization and minimax regret policies persists under these operational constraints, confirming that feasibility requirements rather than estimation uncertainty determine achievable welfare gains in practice. This finding supports implementing welfare maximization policies with modest reassignment limits, as robust alternatives provide minimal additional protection against model misspecification while producing nearly identical assignment decisions.

\begin{table}[ht]
\centering
\caption{Policy Performance With 20\% Reassignment Constraint} 
\label{tab:20pct_welfare}
\begin{tabular}{rrrrr}
  \hline
 & Welfare Max & Frequentist Minimax & Manski Minimax & Pearl Minimax \\ 
  \hline
Welfare & 0.00136 & 0.00127 & 0.00119 & 0.00123 \\ 
  Frequentist Regret & 0.00031 & 0.00014 & 0.00054 & 0.00043 \\ 
  Manski Regret & 0.00051 & 0.00057 & 0.00028 & 0.00052 \\ 
  Pearl Regret & 0.00036 & 0.00038 & 0.00042 & 0.00025 \\ 
   \hline
 \hline
\end{tabular}
\end{table}


\begin{table}[ht]
\centering
\caption{Policy Performance With 10\% Reassignment Constraint} 
\label{tab:10pct_welfare}
\begin{tabular}{rrrrr}
  \hline
 & Welfare Max & Frequentist Minimax & Manski Minimax & Pearl Minimax \\ 
  \hline
Welfare & 0.00096 & 0.00088 & 0.00082 & 0.00096 \\ 
  Frequentist Regret & 0.00030 & 0.00013 & 0.00047 & 0.00030 \\ 
  Manski Regret & 0.00043 & 0.00048 & 0.00023 & 0.00043 \\ 
  Pearl Regret & 0.00000 & 0.00008 & 0.00014 & 0.00000 \\ 
   \hline
 \hline
\end{tabular}
\end{table}


\end{appendix}
\end{document}